\documentclass[prd,twocolumn,floats,floatfix,nofootinbib]{revtex4}
\usepackage[dvips]{graphicx}
\usepackage{graphics}
\usepackage{dcolumn}
\usepackage{amssymb}
\usepackage{bm}
\usepackage{amsmath}
\usepackage{color}
\usepackage{epstopdf}
\usepackage{float}

\def\spose#1{\hbox to 0pt{#1\hss}}

\def\lta{\mathrel{\spose{\lower 3pt\hbox{$\mathchar"218$}}
     \raise 2.0pt\hbox{$\mathchar"13C$}}}
\def\gta{\mathrel{\spose{\lower 3pt\hbox{$\mathchar"218$}}
     \raise 2.0pt\hbox{$\mathchar"13E$}}} 
\newcommand{\be}{\begin{equation}}
\newcommand{\en}{\end{equation}}
\newcommand{\bea}{\begin{eqnarray}}
\newcommand{\ena}{\end{eqnarray}}

\begin{document}

\title{General Boosted Kerr Black Holes and Papapetrou Electrodynamics}

\author{Rafael F. Aranha$^{a}$\footnote{rafael.aranha@uerj.br}, Carlos E. Cede\~no M.$^{b}$\footnote{cecedeno@cbpf.br}, Rodrigo Maier$^{a}$\footnote{rodrigo.maier@uerj.br}, Ivano Dami\~ao Soares$^{b}$\footnote{ivano@cbpf.br}
\vspace{0.5cm}}

\affiliation{$^a$Departamento de F\'isica Te\'orica, Instituto de F\'isica, Universidade do Estado do Rio de Janeiro,\\
Rua S\~ao Francisco Xavier 524, Maracan\~a,\\
CEP20550-900, Rio de Janeiro, Brazil\\
\\
$^b$Centro Brasileiro de Pesquisas F\'{\i}sicas, \\ Rua Dr. Xavier Sigaud, 150, Urca,
CEP 22290-180, Rio de Janeiro, Brazil}

\date{\today}

\begin{abstract}
In this paper the spacetime of a general boosted Kerr black hole relative to a Lorentz
frame at future null infinity is regarded as a background to examine its respective Papapetrou fields. Taking into account its sole Killing vector we evaluate the electric and magnetic components -- in Bondi-Sachs (BS) and Kerr-Schild (KS) coordinates -- of the Maxwell field which comes from spacetime isometries. To this end we consider 
a general timelike observer in KS coordinates. Satisfying sufficient conditions for horizon and ergosphere formation  
it is shown that nonsingular magnetic field configurations are formed around the boosted Kerr black hole while the electric counterpart vanishes. Different magnetic field patterns are discussed in the case of variations of the boost parameter $\gamma$ and direction $\boldsymbol{\hat{n}}$.    
\end{abstract}
\maketitle
\section{Introduction}
The recent direct observations of the gravitational wave emission from binary black holes mergers by the LIGO Scientific Collaboration   and the Virgo Collaboration\cite{1,2,3,4}  established that the initial black holes of each binary had mass ratios $\alpha '$s ranging from  $\alpha \simeq 0.8$  to  $\alpha \simeq 0.53$.  In this sense, the remnant black hole description\cite{ids3} must contain additional parameters –- boost parameters –- connected with its motion relative to the frame of observation. 

\par 
The boosted black hole solution can actually be a natural set for astrophysical processes related to the deformation of the ergosphere and to the electrodynamics effects that result from the rotating black hole moving at relativistic speeds in a direction coinciding or not with the rotation axis. This is the case of an astrophysical configuration in which an external gas can support electric currents that create large-scale magnetic fields. Motion of black holes in this externally supplied magnetic field can then lead to an electromagnetic  extraction of energy.
\indent \par Two distinct possibilities for the electromagnetic extraction of energy from spiraling black holes are in case. First two orbiting  black holes possesses nonzero angular momentum which induces rotation of spacetimes. Rotating space times can generate electromagnetic outflows, in a manner similar to the classical Faraday disk. This is the physics behind the Blandford-Znajek process\cite{blandford} of extracting the rotational power of  a black hole. In the second, Morozova et al. \cite{morozova} derived analytic solutions of the Maxwell equations for a rotating black hole moving at constant speed in an asymptotically uniform magnetic test field; electromagnetic energy losses computed from accelerated charged particles along the magnetic field lines, that approximate numerical relativity calculations in a force-free magnetosphere.
\indent 
\par Within this context, we will follow the analysis of a system composed of a boosted black hole and the behavior of magnetic fields around them. In \cite{aranha}, we studied the magnetic field generated by the isometries of a boosted rotating black hole given a specific direction in space, where the boost direction coincided with the black hole rotation axis. In this paper, we extend our analysis considering a boost in an arbitrary direction, thus losing the axial symmetry as seen in the previous work. Such a loss of symmetry introduces a greater richness of results as we will see later.
\indent \par We organize the paper as follows: in Section II we build the boosted Kerr solution for an arbitrary direction. %eliminating the symmetry associated with the axial Killing vector. 
From the metric obtained in \cite{ids1}, in Robinson-Trautman coordinates\cite{rt}, we generate the metric system in both Bondi-Sachs\cite{bondi, sachs} and Kerr-Schild coordinates\cite{kerr, schild}, given a series expansion to the second order of the inverse of the radial coordinate which characterizes the observer at infinity. We also indicate the components of the Arnowitt-Deser-Misner\cite{adm} variables for the metric written in Kerr-Schild coordinates. In Section III, the causal structure of the black hole is studied, as well as its ergosphere. In this sense, the different event horizons, as well as the ergosphere, are described in the different coordinate systems as described above. In Section IV, we construct the electromagnetic fields generated by the Papapetrou field\cite{papapetrou} (in this case for timelike Killing vector, TLKV) for an observer at infinity. With these fields in hand, we built several plots which describe the behavior of the fields for different values of the black hole parameters -- boost velocity, boost direction and rotation.
Finally, in Section V, we discuss the final remarks of this work indicating possible generalizations and astrophysical applications.

\section{The Kerr Geometry for a General Boost}
In order to obtain the Kerr metric for a general boost we will follow the same procedure adopted in the axisymmetric case\cite{ids1}. 
We start by considering the Kerr metric\cite{ids2} in Robinson-Trautman coordinates $(u, r, \theta,\phi)$
\begin{eqnarray}
\nonumber
ds^2=\frac{r^2+\Sigma^2(\theta)}{K^2(\theta)}(d\theta^2+\sin^2\theta d\phi^2)~~~~~~~~~~~~~~~~~~\\
\label{igrg}
-2\Big[du+\frac{\omega_0 \sin^2\theta}{K^2(\theta)}d\phi\Big]\Big[dr-\frac{\omega_0\sin^2\theta}{K^2(\theta)}d\phi\Big]~~~~~~~~~\\
\nonumber
-\Big[du+\frac{\omega_0 \sin^2\theta}{K^2(\theta)}d\phi\Big]^2\Big[\frac{r^2-2m_0 r+\Sigma^2(\theta)}{r^2+\Sigma^2(\theta)}\Big],
\end{eqnarray}
where
\begin{eqnarray}
\nonumber
K(\theta)&=&a+b\cos \theta~~{\rm with}~~a=\cosh\gamma,~~b=\sinh\gamma ~~{\rm and}\\
\nonumber
\Sigma(\theta)&=&\frac{\omega_0}{K(\theta)}(b+a\cos\theta).
\end{eqnarray}
In the above, $\omega_0$ and $m_0$ are respectively connected to the rotation and mass parameters of the exact solution (\ref{igrg}) of Einstein field equations. The Robinson-Trautman coordinates are related to the Bondi-Sachs asymptotic configuration $(U, R, \theta, \phi)$ through the following transformations\cite{rit} 
\begin{eqnarray}
\label{rit1}
r=K(\theta) R,~~ dr=K(\theta) dR, ~~du=\frac{dU}{K(\theta)}.
\end{eqnarray}
Moreover, a careful analysis indicates\cite{ids1} that the rotation and mass parameters
should transform as 
\begin{eqnarray}
\label{tp}
\omega_0= \omega_b(\theta) K(\theta),~~m_0=m_b(\theta)K^3(\theta).
\end{eqnarray}
Using (\ref{rit1}) and (\ref{tp}) in (\ref{igrg}) the boosted metric of \cite{ids2} asymptotically assumes the form in the Bondi-Sachs coordinates:
\begin{eqnarray}
\nonumber
ds^2=(R^2+\tilde{\Sigma}^2)(d\theta^2 + \sin^2\theta d\phi^2)~~~~~~~~~~~~~~~~~~~~~~\\
\label{ireply}
-2(dU+\omega_b \sin^2\theta d\phi)\Big(dR-\frac{\omega_b\sin^2\theta}{K^2}d\phi\Big)~~~~~~~~~~~~\\
\nonumber
-(dU+\omega_b \sin^2\theta d\phi)^2\Big(\frac{1}{K^2}-\frac{2m_b R}{R^2+\tilde{\Sigma}^2}\Big)
+\mathcal{O}\Big(\frac{1}{R^2}\Big),
\end{eqnarray}
where $\tilde{\Sigma}\equiv\Sigma/K$.

In this same vein, let us now consider the extension of (\ref{igrg}) according to \cite{ids3}. In this more general case
the line element in Robinson-Trautman coordinates is given by
\begin{eqnarray}
\nonumber
ds^2=\frac{r^2+\Sigma^2(\theta, \phi)}{K^2(\theta, \phi)}(d\theta^2+\sin^2{\theta}d\phi^2)~~~~~~~~~~~~~~~~~~~~\\
\nonumber
-2[du-2{\cal L}(\theta, \phi)\cot{(\theta/2)}d\phi]~~~~~~~~~~~~~~~~~~~~~~\\
\label{e1}
\times \Big\{dr+ \omega_0\Big[\frac{n_2\sin{\phi}-n_3\cos{\phi}}{K^2(\theta, \phi)}\Big]d\theta~~~~~~~~~
\\
\nonumber
+\omega_0\Big[\frac{(n_2\cos{\phi}+n_3\sin{\phi})\sin\theta\cos\theta-n_1 \sin^2{\theta}}{K^2(\theta, \phi)}\Big]d\phi \Big\} \\
\nonumber
- [du-2{\cal L}(\theta, \phi)\cot{(\theta/2)}d\phi]^2~~~~~~~~~~\\
\nonumber
\times
\Big[\frac{r^2-2m_0 r+\Sigma^2(\theta, \phi)}{r^2+\Sigma^2(\Theta, \phi)}\Big],
\end{eqnarray}
where the function ${\cal L}(\theta, \phi)$ satisfies the differential equation
\begin{eqnarray}
\label{eqL}
{\cal L}_{,\theta}-\frac{{\cal L}}{\sin\theta}+(1-\cos\theta)\frac{\Sigma(\theta, \phi)}{K^2(\theta, \phi)}=0.
\end{eqnarray}
In the above, 
\begin{eqnarray}
\label{e2}
K(\theta, \phi)&=&a+b(\boldsymbol{\hat{x}}\cdot\boldsymbol {\hat{n}}),\\
\Sigma(\theta, \phi)&=&\frac{\omega_0}{K(\theta, \phi)}[b+a({\boldsymbol{\hat{x}}}\cdot\boldsymbol {\hat{n}})].
\end{eqnarray}
Here ${\boldsymbol {\hat{x}}}\equiv (\cos\theta, \sin\theta\cos \phi,\sin\theta\sin\phi)$ is the unit vector
along an arbitrary direction and $\boldsymbol{\hat{n}} = (n_1, n_2, n_3)$ 
is the boost direction satisfying
\begin{eqnarray}
n_1^2+n_2^2+n_3^2=1.
\end{eqnarray}
It can be shown that the solution of the differential equation \eqref{eqL} is given by
\begin{eqnarray}
\nonumber
{\cal L}(\theta, \phi)=-(1-\cos \theta ) \csc \theta \Big[(\Delta_1 \Delta_7 + \Delta_6 n_1) \omega_0  \cosh \gamma \\
\nonumber
+\frac{\omega_0}{2}( \Delta_8   \sinh \gamma - \text{csch}^2\gamma)\Big],
\end{eqnarray}
where
\begin{widetext}
\begin{align}
\nonumber
	\Delta_1&=n_2 \cos \phi+ n_3 \sin \phi, \\
\nonumber	
    \Delta_2&=\sinh ^2\gamma \left(\Delta_1^2+n_1^2\right)-\cosh ^2\gamma,\\
\nonumber    
    \Delta_3&= \cosh \gamma -n_1 \sinh \gamma,  \\
\nonumber    
    \Delta_4&= \Delta_1 \sinh \gamma  \sin \theta +\cosh \gamma + n_1 \sinh \gamma \cos \theta,\\
\nonumber    
    \Delta_5&= \left(\Delta_1^2+n_1^2\right) \left[2 n_1^2 \sinh ^3\gamma %\left(\Delta_1^2+n_1^2\right) 
    \left(\sin \theta \left(4 \Delta_1^2+n_1^2\right)+3 \Delta_1 n_1 \cos \theta \right) %\nonumber\\
    %& 
    -2 \sinh \gamma  \cosh ^2\gamma %\left(\Delta_1^2+n_1^2\right) 
    \left(\sin \theta \left(8 \Delta_1^2+ n_1^2\right)+5 \Delta_1 n_1 \cos \theta\right) \right]\nonumber\\
    & +\cosh ^3 \gamma  \left(2 \Delta_1 \coth ^2\gamma -7 \Delta_1 \left( \Delta_1^2+n_1^2\right)+4 \Delta_1 \coth \gamma  ( \Delta_1
    \sin \theta + n_1 \cos \theta )-( n_1^2- \Delta_1^2) (n_1 \sin (2 \theta )- \Delta_1 \cos (2 \theta ))\right)\nonumber\\
\nonumber    
    &+ \sinh ^2\gamma  \cosh \gamma  \left( \Delta_1^2+ n_1^2\right) \left(-4 \Delta_1^3+\left(n_1^2-4 \Delta_1^2\right) (n_1 
    \sin(2 \theta )- \Delta_1 \cos (2 \theta )) + 5 \Delta_1 n_1^2\right), 
\end{align}
\begin{align}    
    \Delta_6&= \frac{\sinh ^3\gamma  \left(\Delta_1 \sin \theta \left(n_1^2-2 \Delta_1^2\right)+n_1
    	\cos \theta  \left(2 n_1^2-\Delta_1^2\right)\right)}{2 \Delta_2^2 \Delta_4^2} \nonumber \\
    &+\frac{\cosh ^3\gamma  \left(-\Delta_1^2 \cos (2 \theta )-2 \left( \Delta_1^2+ n_1^2\right)+\coth \gamma  (\coth \gamma +2  \Delta_1 \sin \theta +2 n_1 \cos \theta )+ \Delta_1  n_1 \sin (2 \theta )\right)}{2 \Delta_2^2 \Delta_4^2 \left( \Delta_1^2+ n_1^2\right)}\nonumber\\
    &+\frac{\Delta_1 \sinh ^2\gamma  \cosh \gamma  \left(-7 \Delta_1^3-5
    	\left(\Delta_1^2+ n_1^2\right) ( n_1 \sin (2 \theta )- \Delta_1 \cos (2 \theta ))-5
    	\Delta_1  n_1^2\right)}{4 \Delta_2^2 \Delta_4^2 \left(\Delta_1^2+ n_1^2\right)} \nonumber \\
    & +\frac{n_1^4 \sinh \gamma  \sinh (2 \gamma )}{4 \Delta_2^2 \Delta_4^2 \left( \Delta_1^2+ n_1^2\right)}-\frac{\sinh \gamma  \cosh ^2\gamma  (3 \Delta_1 \sin \theta +2 n_1 \cos \theta )}{\Delta_2^2 \Delta_4^2} %\nonumber\\    
    %& 
\nonumber    
    -\frac{3 \Delta_1 n_1 \sinh ^2\gamma  \tanh ^{-1}\left(\frac{\Delta_1 \sinh \gamma + \Delta_3 \tan \left(\frac{\theta }{2}\right)}{\sqrt{\Delta_2}}\right)}{\Delta_2^{5/2}},\\
\nonumber    
    \Delta_7&= \frac{ \Delta_5}{4 \Delta_2^2 \Delta_4^2 n_1 \left(\Delta_1^2+ n_1^2\right)}-\frac{\left(\cosh ^2\gamma -\sinh ^2\gamma  \left(n_1^2-2 \Delta_1^2\right)\right) \tanh ^{-1}\left(\frac{\Delta_1 \sinh \gamma + \Delta_3 \tan \left(\frac{\theta}{2}\right)}{\sqrt{\Delta_2}}\right)}{\Delta_2^{5/2}},\\
    \Delta_8&= \frac{6 \Delta_1 \sinh \gamma  \cosh \gamma  \tanh ^{-1}\left(\frac{\Delta_1 \sinh \gamma + \Delta_3 \tan \left(\frac{\theta }{2}\right)}{\sqrt{\Delta_2}}\right)}{\Delta_2^{5/2}} 
    + \frac{ \Delta_1 \left(3 \Delta_1 \sinh \gamma \cosh \gamma +\sin \theta 
    	\left(\cosh ^2\gamma +2 \sinh ^2\gamma  \left(\Delta_1^2+ n_1^2\right)\right)\right)}{\Delta_2^2 \Delta_4 n_1}\nonumber\\
\nonumber    	
    &-\frac{\text{csch} \gamma  \left( \Delta_1 \sinh \gamma  \cosh \gamma 
    	\sin \theta +\cosh ^2\gamma - n_1^2 \sinh ^2 \gamma  \right)}{ \Delta_2  \Delta_4^2 n_1}. 
\end{align}
\end{widetext}
It is easy to show that in the case $\boldsymbol {\hat{n}}=(1, 0, 0)$ the above result reduces to
\begin{eqnarray}
\label{L}
{\cal L}=-\frac{\omega_0\sin^2\theta \tan{(\theta/2)}}{2K^2(\theta)}
\end{eqnarray}
and the exact Kerr solution (\ref{igrg}) may be recovered
so that the axisymmetric configuration (\ref{ireply}) can be reobtained. 

In order to seek for a generalization of axisymmetric configurations we now consider the
extended transformations
\begin{eqnarray}
\label{rit1n}
r=K(\theta, \phi) R,~~ dr=K(\theta, \phi) dR, ~~du=\frac{dU}{K(\theta, \phi)},\\
\label{rit2n}
\omega_0= \omega_b(\theta,\phi) K(\theta,\phi),~~m_0=m_b(\theta,\phi)K^3(\theta, \phi),~~~~
\end{eqnarray}
so that the boosted metric of \cite{ids3} asymptotically assumes the following form in the Bondi-Sachs coordinates:
\begin{widetext}
\begin{eqnarray}
\nonumber
ds^2=\Big[R^2+\frac{\Sigma^2(\theta, \phi)}{K^2(\theta, \phi)}\Big](d\theta^2+\sin^2\theta d\phi^2)-2[dU-2{\cal L}K(\theta, \phi)\cot(\theta/2)d\phi]\Big\{dR+\frac{\omega_b(\theta, \phi)}{K^2(\theta, \phi)}\\
\label{ssbs}
\times\Big[\Big((n_2\cos\phi+n_3\sin\phi) \cos\theta\sin\theta
-n_1\sin^2\theta \Big)d\phi+(n_2\sin\phi -n_3\cos\phi) d\theta\Big]\Big\}
\\
-[dU-2{\cal L}K(\theta, \phi)\cot(\theta/2)d\phi]^2
\nonumber
\Big[\frac{1}{K^2(\theta, \phi)}-\frac{2m_b(\theta,\phi)RK^2(\theta,\phi)}{R^2K^2(\theta,\phi)+\Sigma^2(\theta, \phi)}   \Big]+\mathcal{O}\Big(\frac{1}{R^2}\Big).
\end{eqnarray}
\end{widetext}
It is easy to see that the leading components of the Einstein tensor of (\ref{ssbs}) are of the order $\mathcal{O}(R^{-2})$ or higher so that to an asymptotic Lorentz frame at future null infinity $G_{\mu\nu}\simeq 0$. Applying the coordinate transformations (\ref{rit1n}) in (\ref{ssbs}) we finally obtain the boosted metric (\ref{ssbs}) in Robinson-Trautman coordinates
\begin{widetext}
\begin{eqnarray}
\nonumber
ds^2=-\Big[1-\frac{2r K^3(\theta, \phi)m_b(\theta, \phi)}{r^2+\Sigma^2(\theta, \phi)}\Big]du^2 - 2 du dr
-\frac{2\omega_b(\theta, \phi)}{K(\theta, \phi)}(n_2\sin\phi-n_3\cos\phi)dud\theta~~~~~~~~~~~~~~~~~~~~~~~~~~~~\\
\nonumber
+2\Big\{\frac{\omega_b(\theta, \phi)\sin\theta}{K(\theta, \phi)}[n_1\sin\theta-\cos\theta(n_2\cos\phi+n_3\sin\phi)]+{2}{\cal L}\cot(\theta/2)\Big[1-\frac{2r K^3(\theta, \phi)m_b(\theta, \phi)}{r^2+\Sigma^2(\theta, \phi)}    \Big]\Big\}dud\phi\\
\label{ssrt}
+4\cot(\theta/2){\cal L}dr d\phi + \frac{r^2+\Sigma^2(\theta, \phi)}{K^2(\theta, \phi)}d\theta^2+\frac{4\omega_b(\theta, \phi)\cot(\theta/2){\cal L}}{K(\theta, \phi)}(n_2\sin\phi-n_3\cos\phi)d\theta d\phi~~~~~\\
\nonumber
+\Big\{\frac{\sin^2\theta}{K^2(\theta, \phi)}[r^2+\Sigma^2(\theta, \phi)]-\frac{4\cot(\theta/2){\cal L}\omega_b(\theta, \phi)}{K(\theta, \phi)}[n_1\sin^2\theta -(n_2\cos\phi +n_3\sin\phi)\cos\theta\sin\theta]~~~~~~\\
\nonumber
-4\cot^2(\theta/2){\cal L}^2\Big[1-\frac{2rK^3(\theta, \phi) m_b(\theta, \phi)}{r^2+\Sigma^2(\theta, \phi)}\Big]\Big\}d\phi^2
+\mathcal{O}\Big(\frac{1}{r^2}\Big).
\end{eqnarray}
\end{widetext}
To complete our task we now apply the transformation $u = t-r$ in order to bring (\ref{ssrt}) to the Kerr-Schild form. In this case we end up with
\begin{widetext}
\begin{eqnarray}
\label{ks}
\nonumber
ds^2=-\Big[1-\frac{2r K^3(\theta, \phi) m_b(\theta, \phi)}{r^2+\Sigma^2(\theta, \phi)}\Big]dt^2
-\frac{4 r K^3(\theta, \phi)m_b(\theta, \phi)}{r^2+\Sigma^2(\theta, \phi)}dt dr
-\frac{2\omega_b(\theta, \phi)(n_2\sin\phi-n_3\cos\phi)}{K(\theta, \phi)}dtd\theta\\
\nonumber
+2\Big\{\frac{\omega_b(\theta, \phi)\sin\theta}{K(\theta, \phi)}[n_1\sin\theta-\cos\theta(n_2\cos\phi+n_3\sin\phi)]
+{2}{\cal L}\cot(\theta/2)\Big[1-\frac{2rK^3(\theta, \phi) m_b(\theta, \phi)}{r^2+\Sigma^2(\theta, \phi)}    
\Big]\Big\}dt d\phi\\
\nonumber
+\Big[1+\frac{2rK^3(\theta, \phi) m_b(\theta, \phi)}{r^2+\Sigma^2(\theta, \phi)}\Big]dr^2
+\frac{2\omega_b(\theta, \phi)(n_2\sin\phi-n_3\cos\phi)}{K(\theta, \phi)}drd\theta\\
+{2}\Big\{\frac{\omega_b(\theta, \phi)\sin\theta[(n_2\cos\phi+n_3\sin\phi)\cos\theta-n_1\sin\theta]}{K(\theta, \phi)}
+\frac{4\cot(\theta/2){\cal L}r K^3(\theta,\phi) m_b(\theta, \phi) }{r^2+\Sigma^2(\theta, \phi)}\Big\}dr d\phi\\
\nonumber
+ \frac{r^2+\Sigma^2(\theta, \phi)}{K^2(\theta, \phi)}d\theta^2
+\frac{4\omega_b(\theta, \phi)\cot(\theta/2){\cal L}}{K(\theta, \phi)}(n_2\sin\phi-n_3\cos\phi)d\theta d\phi\\
\nonumber
+\Big\{\frac{\sin^2\theta}{K^2(\theta, \phi)}[r^2+\Sigma^2(\theta, \phi)]-\frac{4\cot(\theta/2){\cal L}\omega_b(\theta, \phi)}{K(\theta, \phi)}[n_1\sin^2\theta -(n_2\cos\phi +n_3\sin\phi)\cos\theta\sin\theta]\\
\nonumber
-4\cot^2(\theta/2){\cal L}^2\Big[1-\frac{2rK^3(\theta, \phi) m_b(\theta, \phi)}{r^2+\Sigma^2(\theta, \phi)}\Big]\Big\}d\phi^2+\mathcal{O}\Big(\frac{1}{r^2}\Big).
\end{eqnarray}
\end{widetext}
For completeness, it is worth to evaluate the lapse function $N$, the shift $N^i$ and the spatial metric $\gamma_{ij}$ of (\ref{ks}). Writing the ADM decomposition of (\ref{ks}) as
\begin{eqnarray}
\label{adm}
ds^2=-N^2 dt^2+\gamma_{ij}(dx^i-N^idt)(dx^j-N^jdt),
\end{eqnarray}
the components of the spatial metric are given by
\begin{widetext}
\begin{eqnarray}
\nonumber
\gamma_{rr}&=&1+\frac{2r K^3(\theta, \phi) m_b(\theta, \phi)}{r^2+\Sigma^2(\theta, \phi)}, ~~\gamma_{r\theta}=\frac{\omega_b(\theta, \phi)}{K(\theta, \phi)}(n_2\sin\phi-n_3\cos\phi),\\
\nonumber
\gamma_{r\phi}&=&\frac{4r K^3(\theta, \phi)\cot(\theta/2)m_b(\theta, \phi){\cal L}}{r^2+\Sigma^2(\theta, \phi)}+\frac{\omega_b(\theta, \phi)\sin\theta}{K(\theta, \phi)}[\cos\theta(n_2\cos\phi+n_3\sin\phi)-n_1\sin\theta],\\
\nonumber
\gamma_{\theta\theta}&=&\frac{r^2+\Sigma^2(\theta, \phi)}{K^2(\theta, \phi)},~~ 
\gamma_{\theta \phi}=\frac{2\cot(\theta/2)\omega_b(\theta, \phi){\cal L}}{K(\theta, \phi)}(n_2\sin\phi-n_3\cos\phi),\\
\nonumber
\gamma_{\phi\phi}&=&\frac{\sin^2\theta[r^2+\Sigma^2(\theta, \phi)]}{2K^2(\theta, \phi)}+\frac{2{\cal L}}{K(\theta, \phi)}
\Big\{2\cos^2(\theta/2)\omega_b(\theta, \phi)[\cos\theta(n_2\cos\phi+n_3\sin\phi)-n_1\sin\theta]\\
\nonumber
&&-\cot^2(\theta/2)K(\theta, \phi)
{\cal L}\Big[1-\frac{2rK^3(\theta, \phi)m_b(\theta, \phi)}{r^2+\Sigma^2(\theta, \phi)}\Big]\Big\}.
\end{eqnarray}
\end{widetext}
The shift components on the other hand read,
\begin{eqnarray}
\nonumber
N^r &=&1-\frac{\gamma_{\theta\theta}\gamma_{\phi\phi}-\gamma^2_{\theta\phi}-2(\gamma_{r\phi}\gamma_{\theta\theta}-
\gamma_{r\theta}\gamma_{\theta\phi})\cot(\theta/2){\cal L}}{\tilde{\gamma}},\\
\nonumber
N^\theta &=&\frac{\gamma_{r\theta}\gamma_{\phi\phi}-\gamma_{r\phi}\gamma_{\theta\phi}+2(\gamma_{rr}\gamma_{\theta\phi}-\gamma_{r\theta}\gamma_{r\phi})\cot(\theta/2){\cal L}}{\tilde{\gamma}},\\
\nonumber
N^\phi &=&-\frac{\gamma_{r\theta}\gamma_{\theta\phi}-\gamma_{r\phi}\gamma_{\theta\theta}-2(\gamma^2_{r\theta}-\gamma_{rr}\gamma_{\theta\theta})\cot(\theta/2){\cal L}}{\tilde{\gamma}},
\end{eqnarray}
where $\tilde{\gamma}\equiv {\rm det}(\gamma_{ij})$. Finally, the lapse function is given by
\begin{eqnarray}
N=\pm\sqrt{1+\gamma_{ij}N^iN^j-\frac{2rK^3(\theta, \phi)m_b(\theta,\phi)}{r^2+\Sigma^2(\theta, \phi)}}.
\end{eqnarray}

Although these last results are not on the scope of the present paper, we remark that they might be 
important in a plethora of configurations -- for instance, to fix a zero angular momentum observer\cite{takahashi}. 

\section{The Causal Structure and the Ergosphere}

In order to define the causal structure of the spacetime (\ref{ssbs}) we turn our attention for the horizon condition in Kerr-Schild coordinates. Fixing $g^{rr}=0$ we obtain that the coordinate singularities $r_\pm$ are given by
\begin{eqnarray}
\label{h1}
r_\pm = K^3(\theta, \phi)\Big[m_b(\theta, \phi)\pm \sqrt{m^2_b(\theta, \phi)-\frac{\omega^2_b(\theta, \phi)}{K^4(\theta, \phi)}}\Big]
\end{eqnarray}
as one should expect\cite{ids3}. Using \eqref{tp} we show that the horizons are spherically symmetric, i.e.,
\begin{eqnarray}
	\label{h2}
	r_\pm = m_0\pm \sqrt{m^2_0 - \omega^2_0 }.
\end{eqnarray} 
The general boosted Kerr black hole also presents ergospheres defined by $g_{tt}|_{r_e}=0$. In Kerr-Schild coordinates this condition turns into
\begin{eqnarray}
\label{ergos}
-\frac{1}{K^2(\theta, \phi)}+\frac{2r_e K(\theta, \phi )m_b(\theta, \phi)}{r_e^2+\Sigma^2(\theta, \phi)}=0,
\end{eqnarray}
and (\ref{ergos}) furnish two solutions, namely,
\begin{eqnarray}
\label{rergos}	
r_{e\pm}= K^3(\theta, \phi)\Big[m_b(\theta, \phi)\pm \sqrt{m^2_b(\theta, \phi)-\frac{\Sigma^2(\theta, \phi)}{K^6(\theta, \phi)}}\Big].
\end{eqnarray}
From the above it is easy to see that in the domain
$
K^2(\theta, \phi) m_b(\theta, \phi) >  w_b(\theta,\phi)$, $
K^3(\theta, \phi)m_b(\theta, \phi)>\Sigma(\theta, \phi)$ and $\Sigma(\theta, \phi)/K(\theta, \phi)<\omega_b(\theta, \phi),$
our solution allows one to obtain an exterior event horizon $r_+$ together with an outer ergosphere at $r_{e+}$. Furthemore, a Cauchy horizon $r_-$ together with an inner ergosphere at $r_{e-}$ is also obtained. It is easy to see that $r_{e-}<r_+$ so that the two relevant structures are basically $r_+$ and $r_{e+}$ (with $r_+<r_{e+}$). In the following we shall restrict ourselves to these configurations.

\begin{figure*}
\centering
\includegraphics[height=5cm]{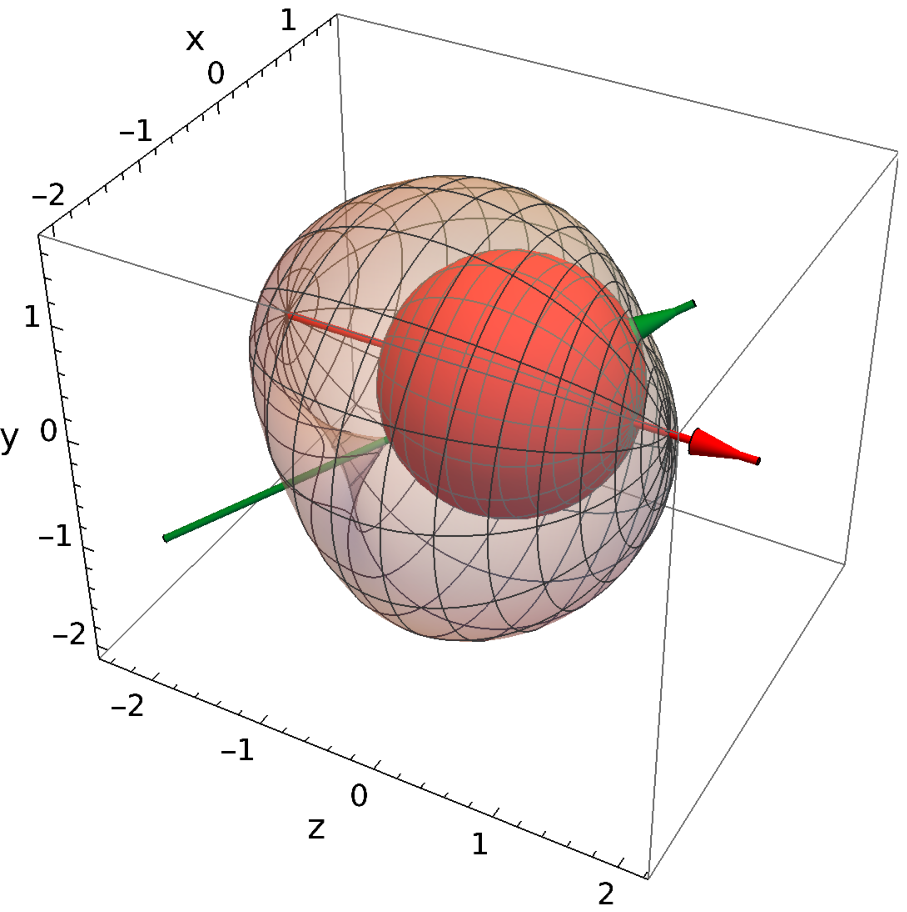}\includegraphics[height=5cm]{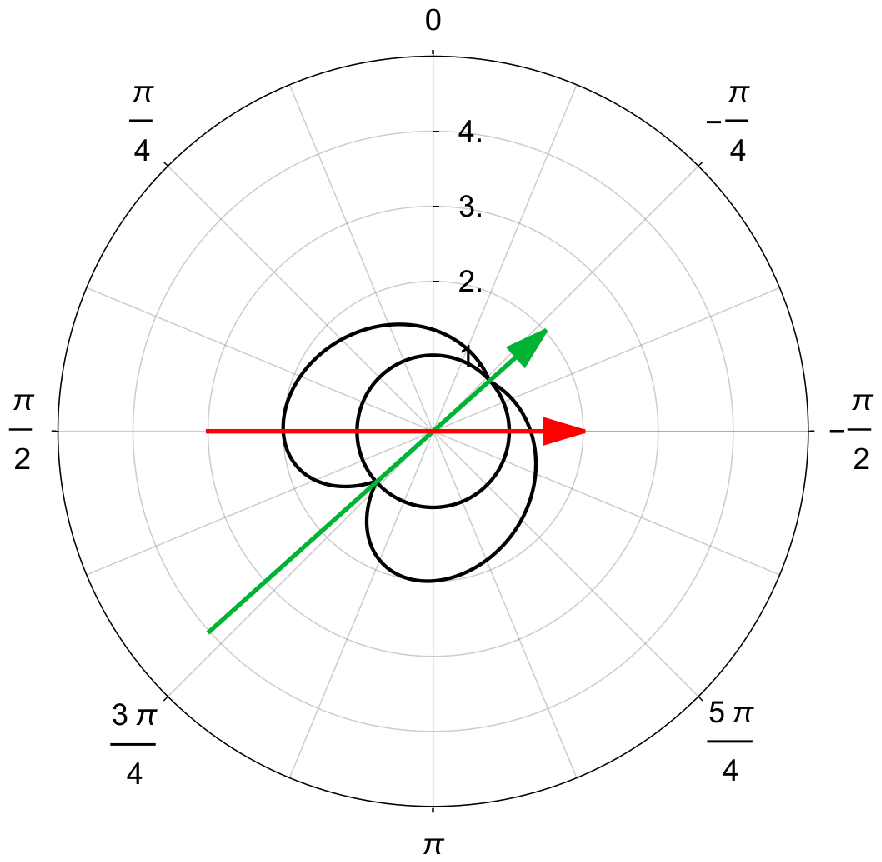}\includegraphics[height=5cm]{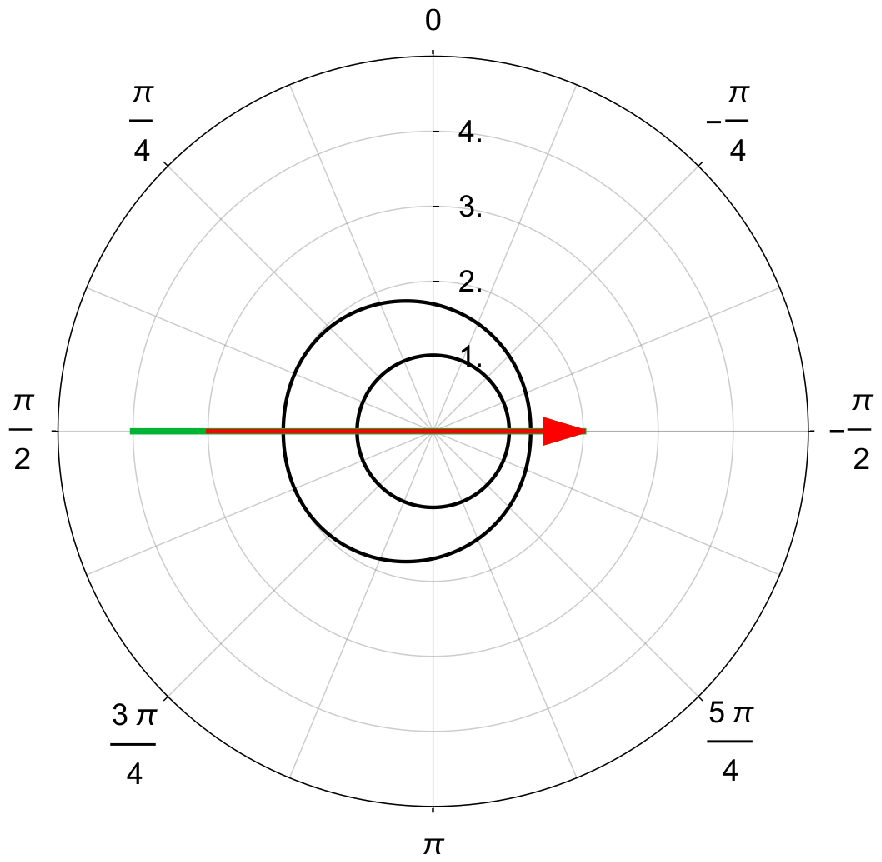}
\caption{
Horizon and ergosphere for a boosted Kerr black hole in the Kerr-Schild coordinates.  Here, $m_0=1.00,\,\omega_0=0.99,\,\gamma=0.90,\,\boldsymbol{\hat{n}}=(0.75,0.50,0.43)$. Left panel: 3D representation, the boost $\boldsymbol{\hat{n}}$ and the rotation axis $\boldsymbol{\hat{z}}$ are represented in green and red respectively. Middle panel: Section along the plane formed by the boost vector $\boldsymbol{\hat{n}}$ and the rotation axis $\boldsymbol{\hat{z}}$. Right panel: Section orthogonal to the last section, containing the rotation axis, through the coordinate origin.
} 
\label{fig1}
\end{figure*}

We write \eqref{rergos}, by using \eqref{tp} and equations \eqref{e2}, as
\begin{align}	
	r_{e\pm}=&m_0 \pm \sqrt{m_0^2-\frac{\omega_0^2 (a \cos \alpha +b)^2}{(a+b \cos \alpha )^2}},
\end{align}
where $\alpha$ is the angle between $\boldsymbol{\hat{x}}$ and $\boldsymbol{\hat{n}}$. It shows us that the ergosphere is axisymmetric with respect to the boost axis $\boldsymbol{\hat{n}}$. In Fig. \ref{fig1} we illustrate the horizon and ergosphere of a boosted Kerr black hole in KS coordinates for a particular configuration.

\begin{figure*}
\centering
\includegraphics[height=5cm]{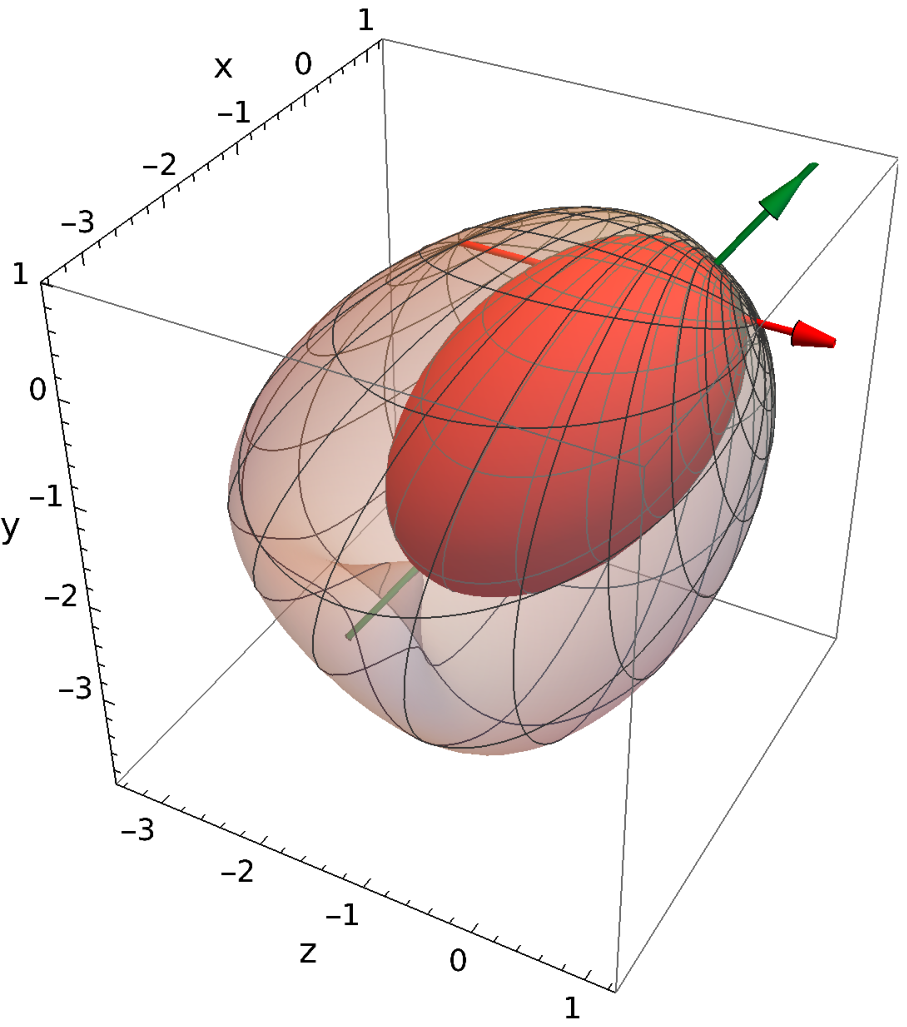}
\includegraphics[height=5cm]{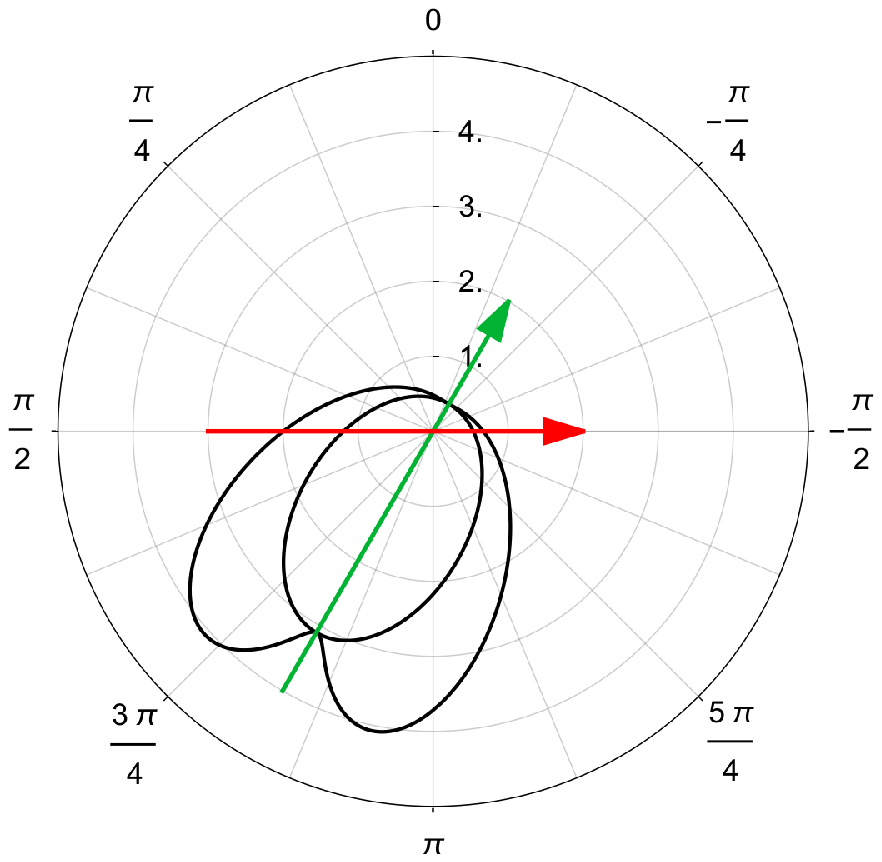}
\includegraphics[height=5cm]{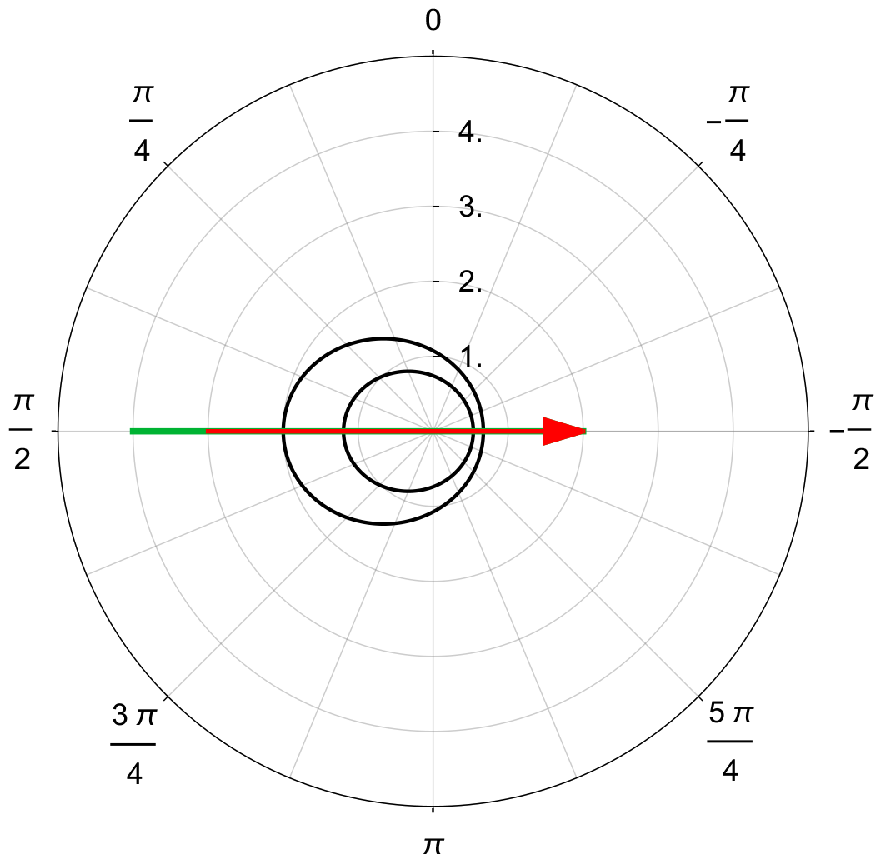}
\caption{Horizon and ergosphere for a boosted Kerr black hole in the Bondi-Sachs coordinates.  Here, $m_0=1.00,\,\omega_0=0.99,\,\gamma=1.30,\,\boldsymbol{\hat{n}}=(0,1,0)$. Left panel: 3D representation, the boost $\boldsymbol{\hat{n}}$ and the rotation axis $\boldsymbol{\hat{z}}$ are represented in green and orange respectively. Middle panel: Section along the plane formed by the boost vector $\boldsymbol{\hat{n}}$ and the rotation axis $\boldsymbol{\hat{z}}$. Right panel: Section orthogonal to the last section, containing the rotation axis, through the coordinate origin.
} 
\label{fig2}
\end{figure*}

In a similar way we examine the horizon and the ergosphere in the Bondi-Sachs coordinates. The horizon $g^{RR}=0$, in these coordinates lead us to surfaces defined by
\begin{eqnarray}
	\label{eq36}
	\nonumber
R_{h\pm}&=& \frac{m_0}{K(\theta ,\phi )}\\
&\pm&\frac{\sqrt{K(\theta ,\phi )^2 \left(m_0^2-\Sigma (\theta ,\phi )^2\right)-\Xi(\theta ,\phi )\, \omega_0
		^2}}{K(\theta ,\phi )^2},~~~~
\end{eqnarray}
where the function $\Xi(\theta,\phi)$ is
\begin{eqnarray}
\nonumber
	\Xi(\theta,\phi)&=& \sin ^2\theta  (-n_1 + n_2 \cot \theta  \cos \phi + n_3 \cot \theta  \sin \phi )^2 \\
	\nonumber
	&&+ ( n_3
	\cos (\phi )- n_2 \sin \phi)^2.
\end{eqnarray}
The horizons $R_{\pm}$ are defined always as
\begin{equation}
	\omega_0^2\leq \frac{m_0^2 K(\theta ,\phi )^2}{\Xi(\theta ,\phi ) \left((a \boldsymbol{\hat{x}}\cdot\boldsymbol{\hat{n}}+b)^2+1\right)}.
\end{equation} 
The ergosphere equation, $g_{UU}=0$, in Bondi-Sachs coordinates reads
\begin{equation}
	\label{eq39}
	R_{e\pm}= \frac{m_0\pm\sqrt{m_0^2-\Sigma (\theta ,\phi )^2}}{K(\theta ,\phi )},
\end{equation}
and these surfaces exist only for the domain
\begin{equation}
	\omega_0 ^2\leq \frac{m_0^2 K(\theta ,\phi )^2}{(a \boldsymbol{\hat{x}}\cdot \boldsymbol{\hat{n}}+b)^2}.
\end{equation}
The explicit form for the ergosphere for a boost in an arbitrary direction $\boldsymbol{n}=(n_1,n_2,n_3)$ with $\|\boldsymbol{\hat{n}}\|=1$, is given by
\begin{eqnarray}
	\label{eq41}
\nonumber	
R_{e+}&=&\frac{m_0}{ a+b \Delta(\theta,\phi) }
\\
&+&\frac{\sqrt{m_0^2 (a+b \Delta(\theta,\phi) )^2-\omega_0 ^2 (a \Delta(\theta,\phi) +b)^2}}{(a+b \Delta(\theta,\phi) )^2}~~~~
\end{eqnarray}
where
\begin{equation}
	\label{eq42}
	\Delta(\theta,\phi)=\boldsymbol{\hat{x}}\cdot \boldsymbol{\hat{n}}=\cos{\alpha}
\end{equation}
with $\alpha$ the angle between $\boldsymbol{\hat{x}}$ and $\boldsymbol{\hat{n}}$. In this form, the ergosphere \eqref{eq41} can be written as
\begin{eqnarray}
\nonumber
	R_{e+}&=&\frac{m_0}{a+b \cos\alpha }\\
	&&+\frac{\sqrt{m_0^2 (a+b \cos\alpha )^2-\omega_0 ^2 (a \cos\alpha +b)^2}}{(a+b \cos\alpha )^2}
\end{eqnarray}
showing that this surface is axially symmetric with respect to the boost axis; moreover this is identical to the axial case, with the condition that $\alpha=\theta$, where $\theta$ is the polar axis. In Fig. \ref{fig2} we illustrate the horizon and ergosphere of a boosted Kerr black hole in BS coordinates for a particular configuration.

It is worth to mention here that, since $\alpha$ is the angle between $\boldsymbol{\hat{x}}$ and $\boldsymbol{\hat{n}}$, the points $\alpha=0$ and $\alpha=\pi$ correspond to the intersections of the ergosphere with the boost axis $\boldsymbol{\hat{n}}$. An evaluation of the $\partial_\alpha R_{e+}$ at this points shows if the ergosphere is smooth as we are expecting. Thus  
\begin{widetext}
\begin{equation}
	\label{derergo}
	\partial_\alpha R_{e+}=\frac{\sin \alpha  \left((a+b \cos \alpha ) \left(\cos \alpha  \left(a^2 \omega_0 ^2-b^2 m_0^2\right)+a b \left(\omega_0
		^2-m_0^2\right)\right)+b m_0 \chi (\alpha ) (a+b \cos \alpha )+2 b \chi (\alpha )^2\right)}{\chi (\alpha ) (a+b \cos \alpha )^3},
\end{equation}
\end{widetext}
where
\begin{equation}
\chi(\alpha)=\sqrt{m_0^2 (a+b \cos\alpha )^2-\omega_0 ^2 (a \cos\alpha +b)^2}.	
\end{equation}
From \eqref{derergo} we observe that 
\begin{equation}
\lim_{\alpha\rightarrow\{0,\pi\}}{\partial_\alpha R_{e+}}=0
\end{equation}
i.e., the ergosphere and at least it's first derivative are continuous.
The exterior ergosphere $R_{e+}$, and the event horizon $R_{h+}$ intersect when $\Xi(\theta,\phi)=0$.
\begin{figure*}[h!]
\centering
\includegraphics[height=5cm]{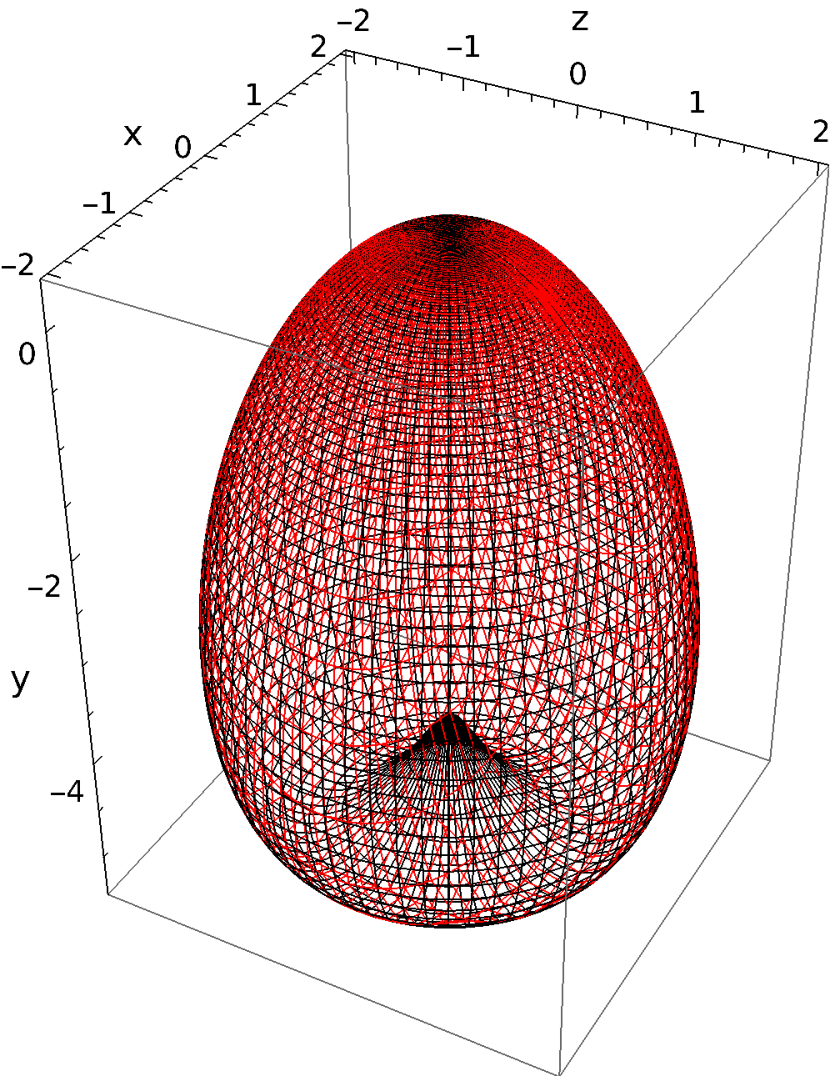}
\includegraphics[height=5cm]{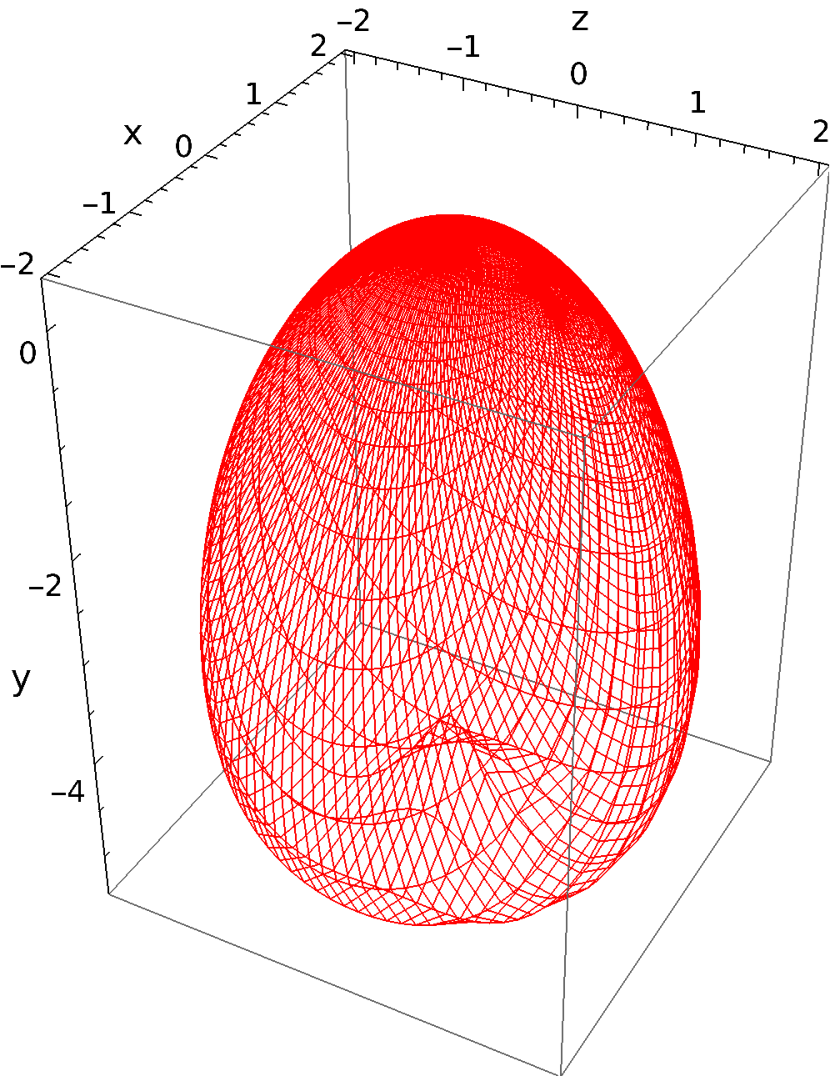}
\includegraphics[height=5cm]{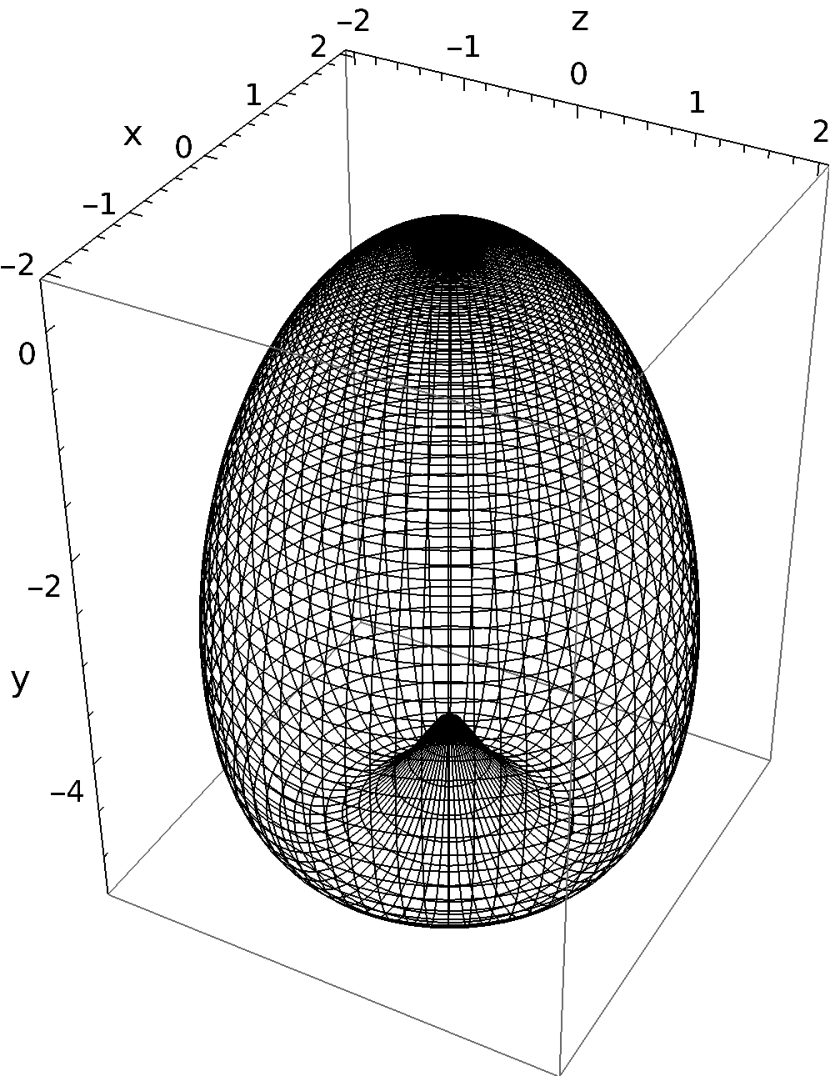}\\
\includegraphics[height=5cm]{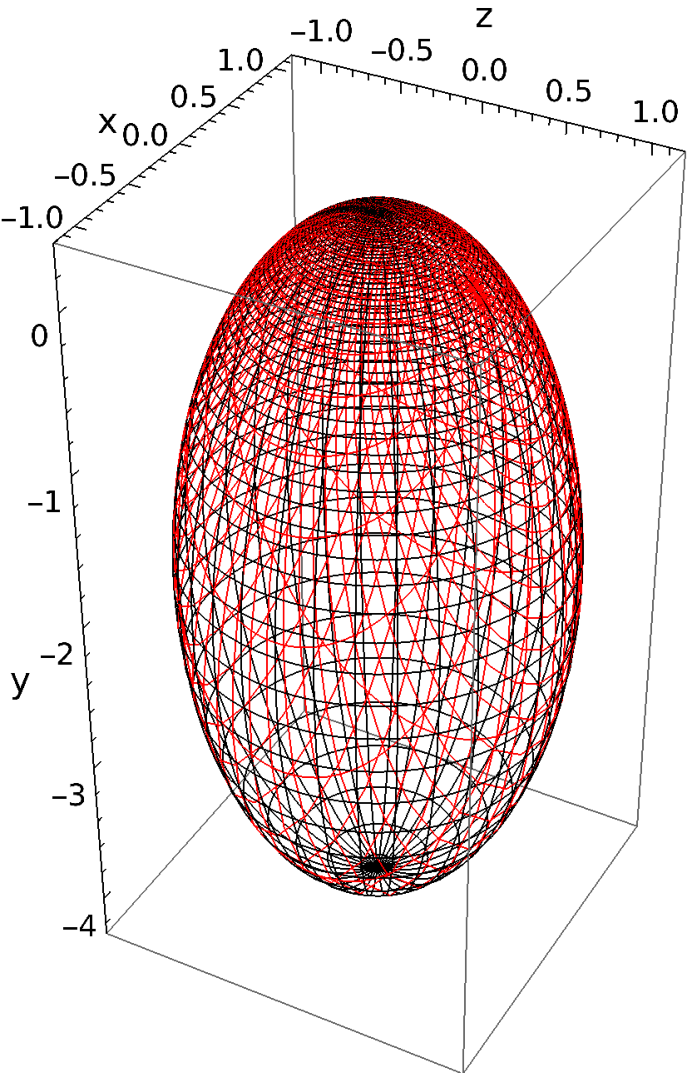}
\includegraphics[height=5cm]{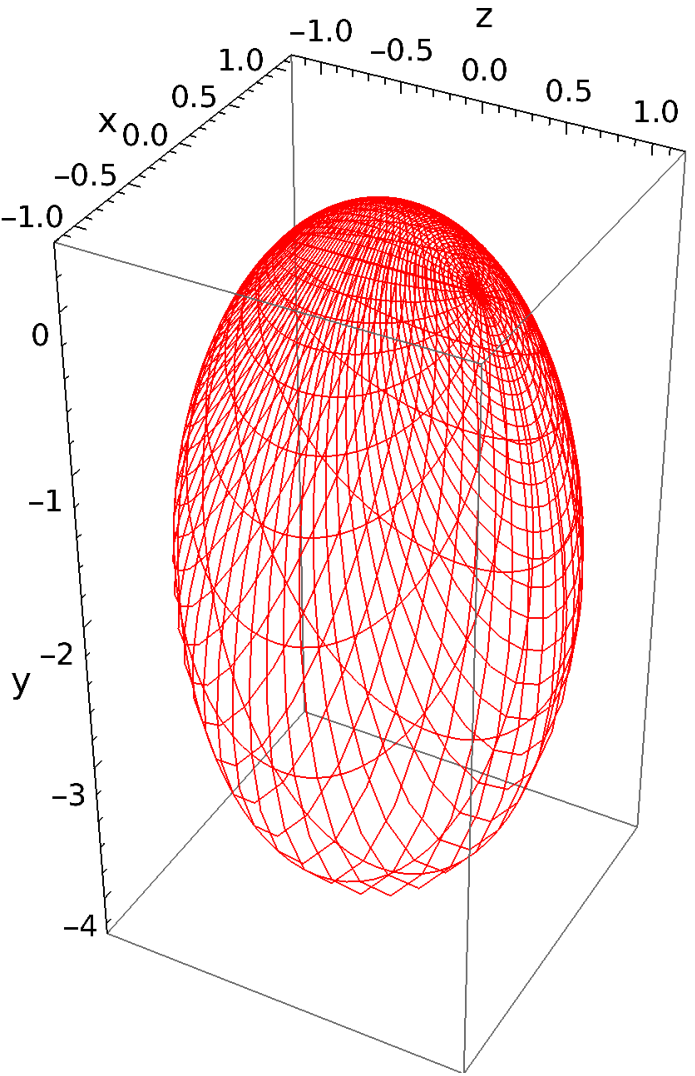}
\includegraphics[height=5cm]{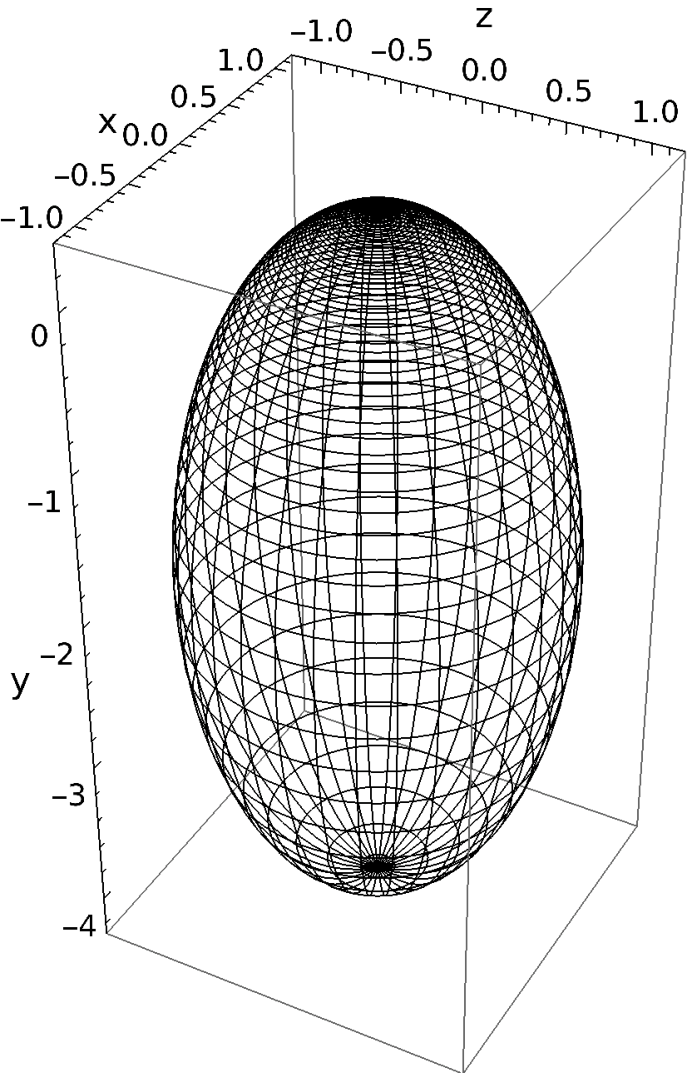}
\caption{Ergosphere and horizon for a boosted Kerr black hole in the Bondi-Sachs coordinates.  Here, $m_0=1.00,\,\omega_0=0.99,\,\gamma=1.30$. Left panel, top and bottom rows: Superposition of two different ergospheres and horizon respectively.  Red and black lines indicate surfaces for $\boldsymbol{\hat{n}}=(0,0,1)$ and $\boldsymbol{\hat{n}}=(1,0,0)$ cases. The ergosphere and the horizon corresponding to $\boldsymbol{n}=(1,0,0)$ were rotated from its original spatial configuration to show the coincidence of both ergospheres and horizon respectively. Middle panel, top and bottom rows: Ergosphere and horizon for $\boldsymbol{\hat{n}}=(0,0,1)$, when the boost axis is orthogonal to the rotation axis $\boldsymbol{z}$. Right panel, top and bottom rows: Ergosphere and horizon for $\boldsymbol{\hat{n}}=(1,0,0)$, where the boost axis $\boldsymbol{\hat{n}}$ and the rotation axis $\boldsymbol{z}$ are parallel. The ergosphere and the horizon were rotated in form that the boost axis now is directed along $\boldsymbol{y}$ direction.
} 
\label{fig3}
\end{figure*}
We also note that the $\Xi(\theta,\phi)$ function can be written just as
\begin{equation}
\Xi(\theta,\phi) = \left(\partial_\theta\Delta (\theta ,\phi )\right)^2+\csc ^2 \theta  \left(\partial_\phi\Delta (\theta ,\phi )\right)^2.
\end{equation}
Taken into account \eqref{eq42}, then
\begin{equation}
	\Xi(\theta,\phi) = \sin^2\alpha.  
\end{equation}
which has two zeros for $\alpha\in [0,\pi]$, they are just for $\alpha=0$ and $\alpha=\pi$, i.e., along the boost axis.
Therefore, the horizons in Eq. \eqref{eq36} correspond to
\begin{eqnarray}
\nonumber
	R_{h\pm}= \frac{m_0 }{a+b \cos\alpha}~~~~~~~~~~~~~~~~~~~~~~~~~~~~~~~~~~~~~~~~~~~~~~ \\
	\pm\frac{\sqrt{m_0^2 (a+b \cos\alpha )^2 -\omega_0 ^2 (a \cos\alpha +b)^2 - \omega_0 ^2\sin^2\alpha  }}{(a+b \cos\alpha )^2},~~
\end{eqnarray}
which shows explicitly axial symmetry with respect to the $\boldsymbol{\hat{n}}$ axis; moreover, in analogy to the ergosphere, this result corresponds to the axisymmetric case, when $\alpha=\theta$. We illustrate this behaviour in Fig. \ref{fig3}.
\section{Basic Electrodynamics from Spacetime Isometry}
It is well known that Killing vectors are solutions of Maxwell equations for vacuum spacetimes\cite{papapetrou}. In fact, let $K_{\mu}$ be a Killing vector which defines the Faraday tensor $F_{\mu\nu}$, namely,
\begin{eqnarray}
\label{e5}
F_{\mu\nu}=\nabla_{\nu}K_{\mu}-\nabla_{\mu}K_{\nu}.
\end{eqnarray}
Therefore, using the Ricci integrability condition $\nabla_{\gamma}\nabla_{\mu}K_{\nu}-\nabla_{\mu}\nabla_{\gamma}K_{\nu}=-R^{\sigma}_{~\nu\mu\gamma}K_{\sigma}$, we obtain
\begin{eqnarray}
\label{e12}
\square K_\gamma = R^\sigma_{~\gamma} K_\sigma,
\end{eqnarray}
and for a vacuum spacetime the equation $\nabla_\mu F^{\mu\nu} =0$ is automatically satisfied.

It can be shown that the above geometry (\ref{ks}) allows one to identify the Killing vector
$\Gamma^\mu=\delta^\mu_{~t}$.
To proceed we now evaluate the Faraday tensor for this Killing vector
in Kerr-Schild coordinates. It turns out that
the sole nonvanishing component of $F_{\mu\nu}$ is given by
\begin{eqnarray}
\label{f23}
{^{t}}F_{23}\simeq\Omega(\theta, \phi)+\frac{4K^3(\theta,\phi) m_b(\theta,\phi)}{r(\cos\theta-1)}\Big({\cal L}-\sin\theta\frac{\partial {\cal L}}{\partial \theta}\Big)\\
\nonumber
+\mathcal{O}\Big(\frac{1}{r^2}\Big).
\end{eqnarray}
where
\begin{widetext}
\begin{eqnarray}
\label{omega}
\nonumber
\Omega(\theta, \phi)=\csc^2\Big(\frac{\theta}{2}\Big)\Big({\cal L}-\sin\theta\frac{\partial {\cal L}}{\partial \theta}\Big)
-\frac{2\omega_b(\theta, \phi)\sin\theta}{K(\theta, \phi)}[n_1\cos\theta+\sin\theta(n_2\cos\phi+n_3\sin\phi)]\\
+\frac{2\omega_b(\theta, \phi)}{K^2(\theta, \phi)}\Big\{(n_2\sin\phi-n_3\cos\phi)\frac{\partial K}{\partial \phi}
+\sin\theta\frac{\partial K}{\partial \theta}[n_1\sin\theta-\cos\theta(n_2\cos\phi+n_3\sin\phi)]\Big\}.
\end{eqnarray}
\end{widetext}
It is easy to see that for ${\boldsymbol{\hat{n}}}=(1, 0, 0)$ we recover the axisymmetric component\footnote{This same component appears
in ref. \cite{aranha} with a typo which must be corrected in order to obtain previous numerical results.}
\begin{eqnarray}
\label{f23ax}
{^{t}}F_{23}\simeq \frac{4 m_b(\theta)\omega_b(\theta)K(\theta)\sin^2\theta}{r}[K^\prime(\theta)-\cot(\theta) K(\theta)]~~~\\
\nonumber
+\mathcal{O}\Big(\frac{1}{r^2}\Big).
\end{eqnarray}
By defining the Hodge dual by ${\cal F}^{\mu\nu}=\frac{1}{2}\epsilon^{\mu\nu\alpha\beta}F_{\alpha\beta}$ from (\ref{f23}) we
obtain
\begin{eqnarray}
\label{f23d}
{^{t}}{\cal F}^{01}\simeq \Omega(\theta, \phi)+\frac{4K^3(\theta,\phi) m_b(\theta,\phi)}{r(\cos\theta-1)}\Big({\cal L}-\sin\theta\frac{\partial {\cal L}}{\partial \theta}\Big)\\
\nonumber
+\mathcal{O}\Big(\frac{1}{r^2}\Big).
\end{eqnarray}

Fixing a proper observer -- connected to the sole Killing vector $\Gamma^\mu$ -- whose $4$-velocity is given by 
$u^{\mu}=(1, 0, 0, 0)$, we are now in a position to evaluate the electric
and magnetic fields using their usual definitions, ${^{t}}{\cal E}^\mu={^{t}}F^{\mu}_{~\nu}u^\nu$ and ${^{t}}{\cal B}^\mu={^{t}}{\cal F}^{\mu}_{~\nu}u^\nu$. It is easy to see that the electric components are negligible -- they fall with powers of $1/r^2$ or higher -- 
while the magnetic counterparts read
\begin{eqnarray}
\label{bs}
{^{t}}{\cal B}^t&\simeq& -\frac{2m_b(\theta, \phi)K^3(\theta, \phi)\Omega(\theta, \phi)}{r} +\mathcal{O}\Big(\frac{1}{r^2}\Big),\\
{^{t}}{\cal B}^r&\simeq& \Omega(\theta, \phi)-\frac{m_b(\theta, \phi)K^3(\theta, \phi)\csc^2(\theta/2)}{r}\\
\nonumber
&\times&\Big[2\Big({\cal L}-\sin\theta
\frac{\partial {\cal L}}{\partial\theta}\Big)+(1-\cos\theta)\Omega(\theta, \phi)\Big]+\mathcal{O}\Big(\frac{1}{r^2}\Big).
\end{eqnarray}
To aid with the interpretation of our results, we sketch in Fig. 4, the behavior of the square of the intensity of the radial magnetic field, $|{^{t}}{\cal B}^r|^2$, when the boosting vector, in green, is directed at an acute angle with respect to the rotation axis $\boldsymbol{\hat{\omega}}$, in red. The most intense regions, for this configuration, are given by two cones between the boost and the rotation direction and by a disk located close to the plane orthogonal to the black-hole spin axis. The spatial distribution of the intensity of the magnetic field, in this and other configurations, are computed in Figs. \ref{FigKSvarGamma}, \ref{FigKSvarBoostDir}, \ref{FigKSvarBoostInten} in Kerr-Schild and Figs. \ref{FigBSvarGamma0}, \ref{FigBSvarw0}, \ref{FigBSvarn} in Bondi-Sachs coordinates, respectively.   
\begin{figure}[h!]
	\includegraphics[height=4cm]{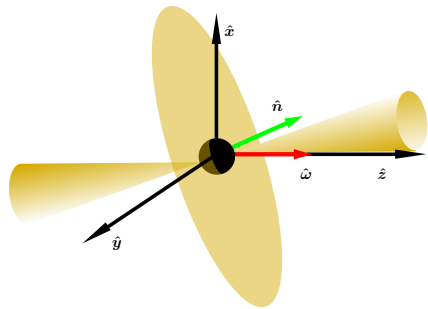}
	\caption{Sketch representing the spatial distribution of the intensity of the radial magnetic field $|{^{t}}{\cal B}^r|^2$, when the boost vector $\boldsymbol{\hat{n}}$ is at acute angle with respect to the rotation axis $\boldsymbol{\hat{\omega}}$.}
	\label{diagram}
\end{figure}
\begin{figure*}%[h!]
\centering
\includegraphics[width=5cm,height=5cm]{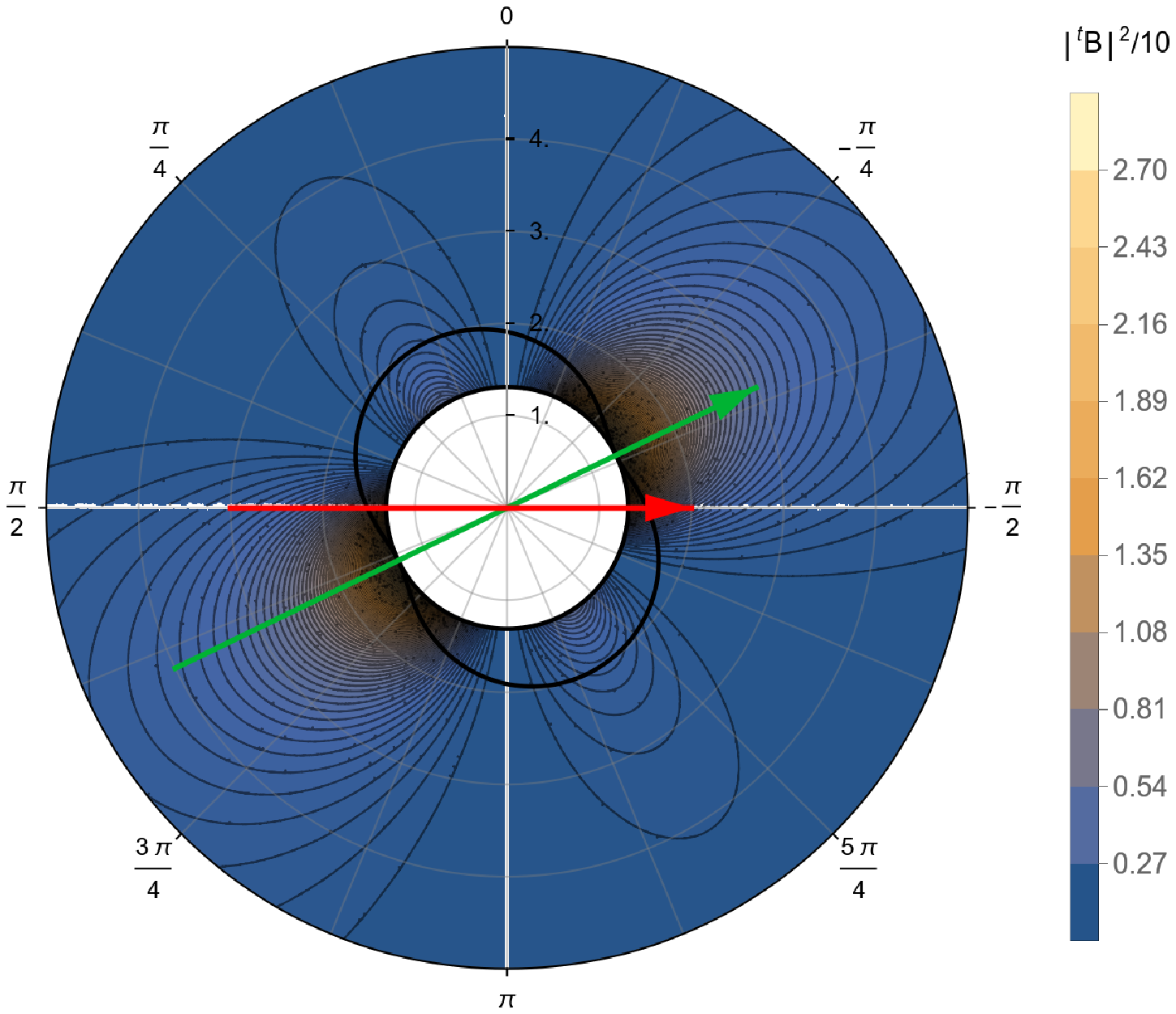}
\includegraphics[width=5cm,height=5cm]{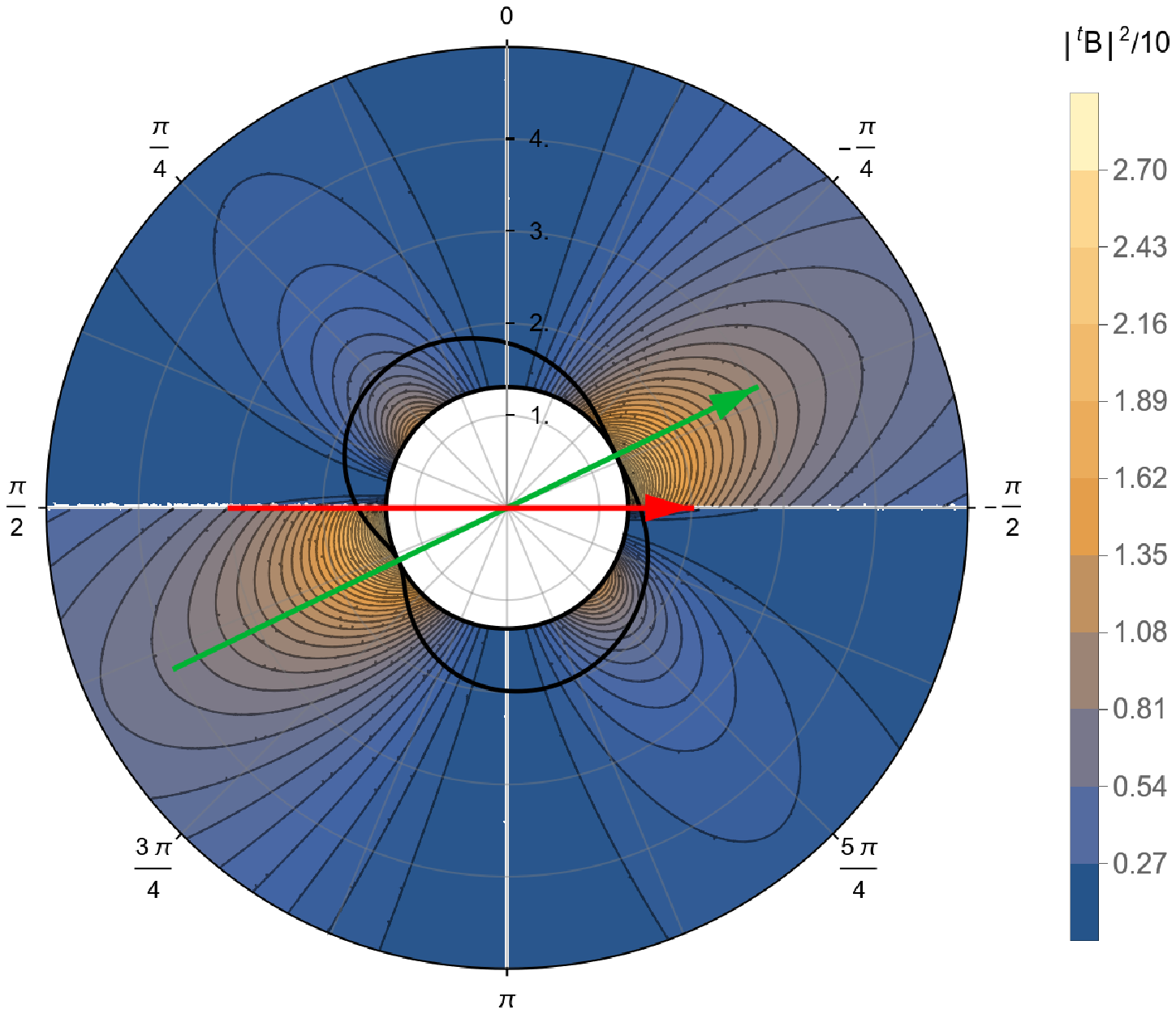}
\includegraphics[width=5cm,height=5cm]{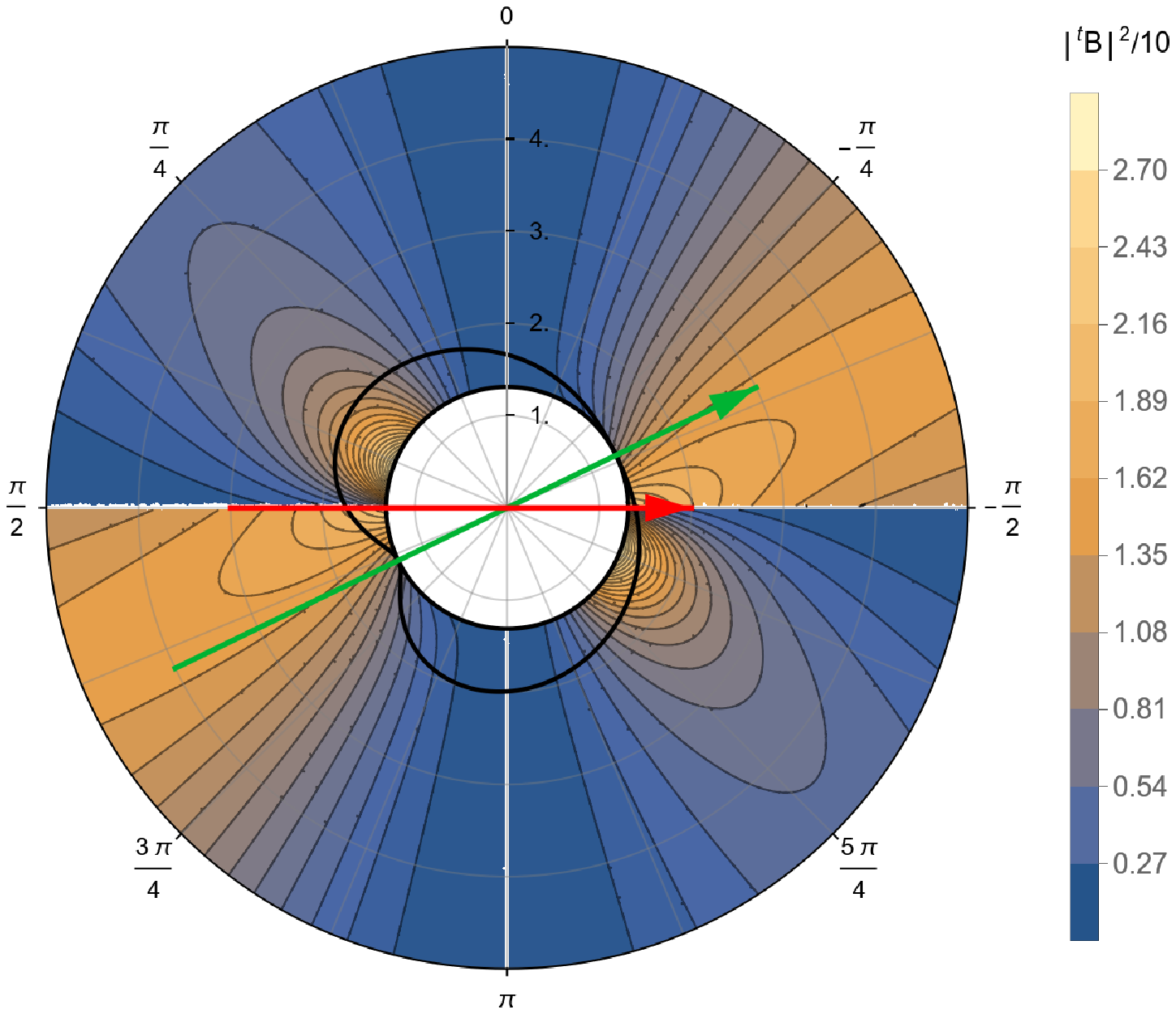}
\caption{Electromagnetic configuration for a Kerr-Schild boosted black-hole for a given boost direction $\boldsymbol{\hat{n}}$ as viewed by an observer with four-velocity $u^{\mu}=(1, 0, 0, 0)$. The green and the red arrows indicate the direction of the boost $\boldsymbol{\hat{n}}$ and the direction of the rotation parameter $\omega$, respectively. Also, $m_0=1, \omega_0=0.95, n_1=0.900, n_2=0.000, n_3=0.436$. We display a plane section containing the boost and the rotation directions in which we show the modulus $|{^{t}}{\cal B}^r|^2$ for different values of the boost parameter $\gamma$. We observe a shear along the rotation axis that increases as $\gamma$ grows. The intensity of the radial magnetic field is higher in the region determined by the boost and the rotation directions. However, close to the normal plane to the boost direction, the magnetic field becomes intense. Left panel: $\gamma=0.0$, Middle panel: $\gamma=0.3$, Right panel: $\gamma=0.6$.
} 
\label{FigKSvarGamma}
\end{figure*}
\begin{figure*}%[h!]
\centering
\includegraphics[height=5cm]{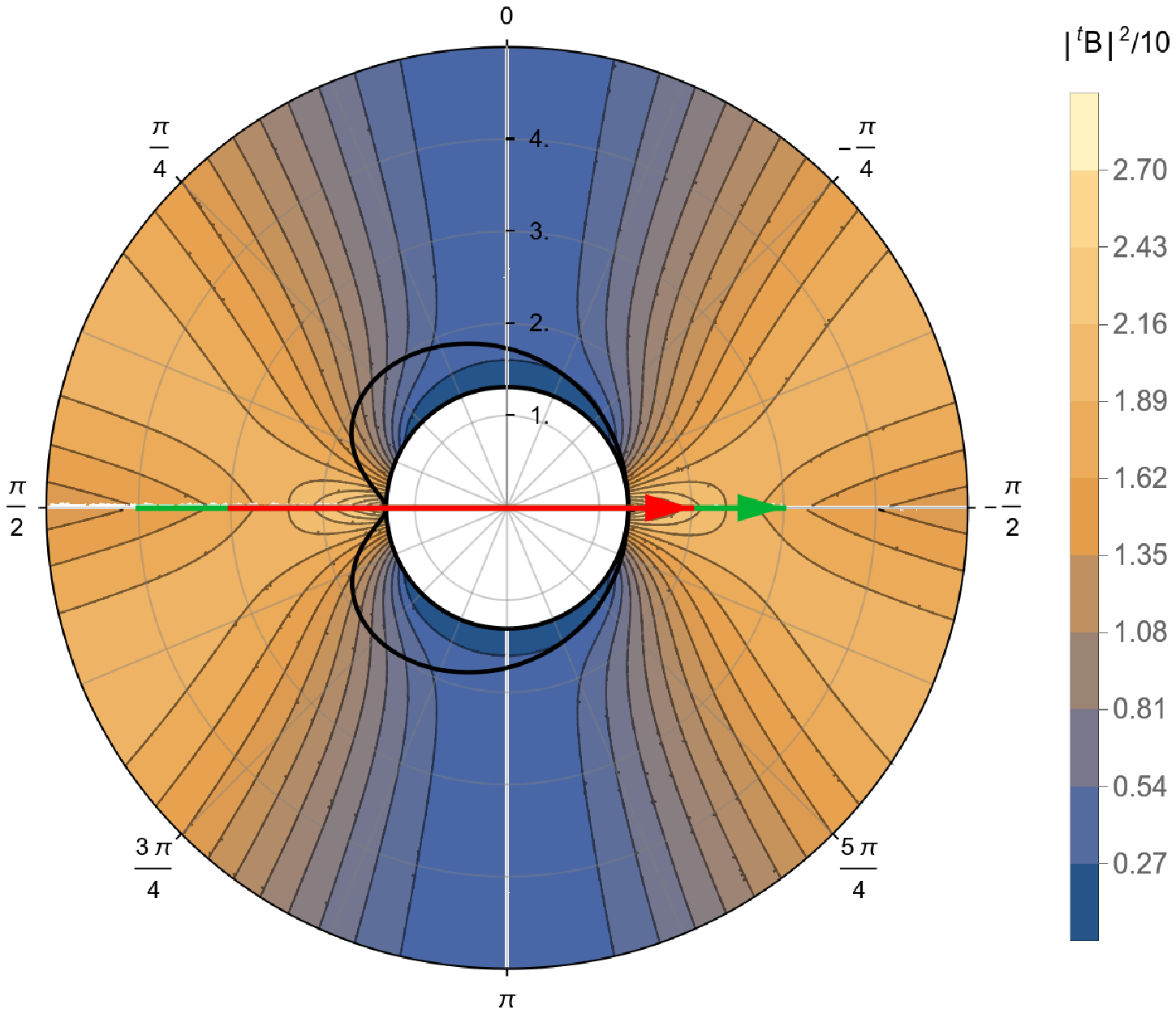}
\includegraphics[height=5cm]{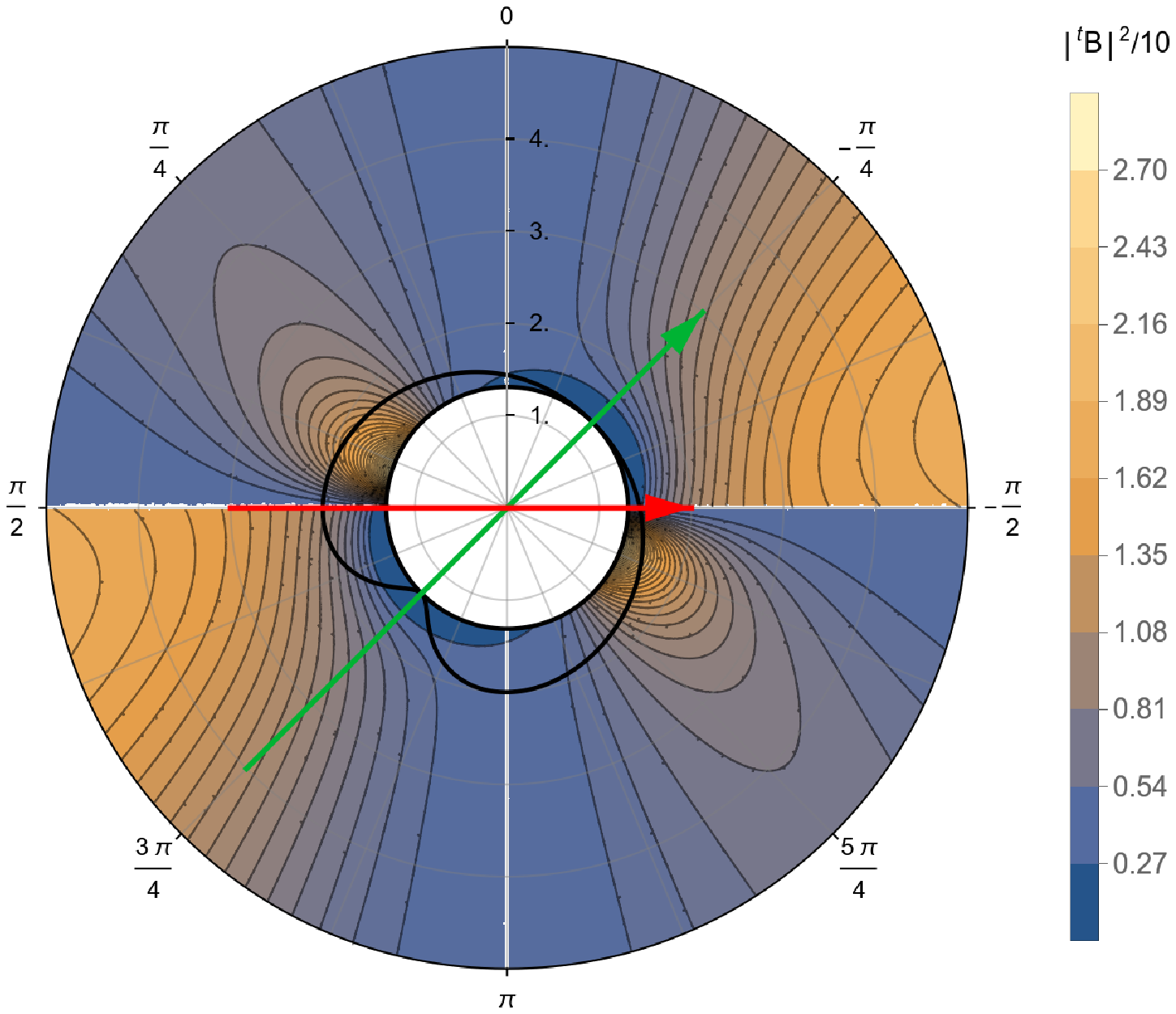}
\includegraphics[height=5cm]{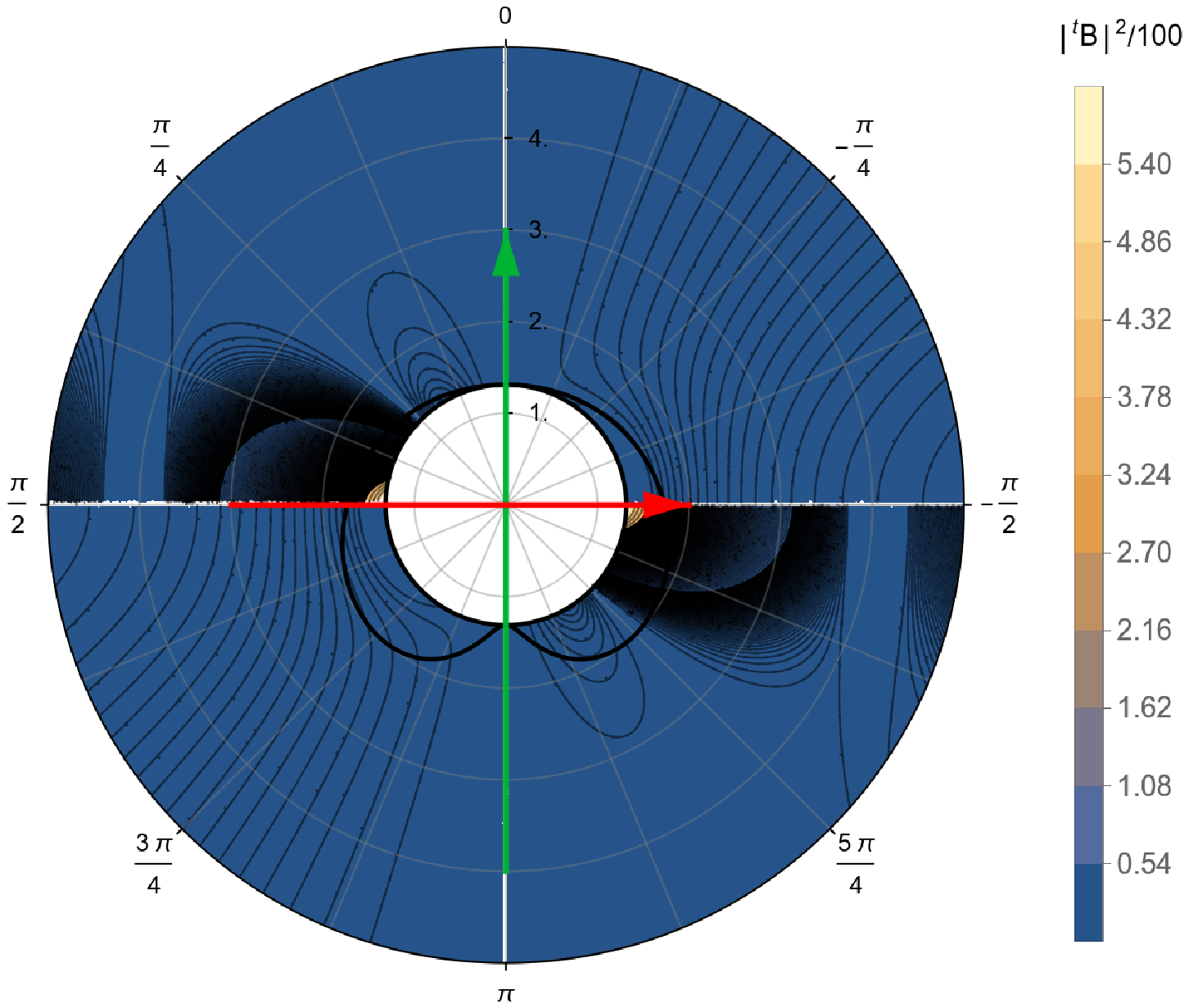}\\
\includegraphics[height=5cm]{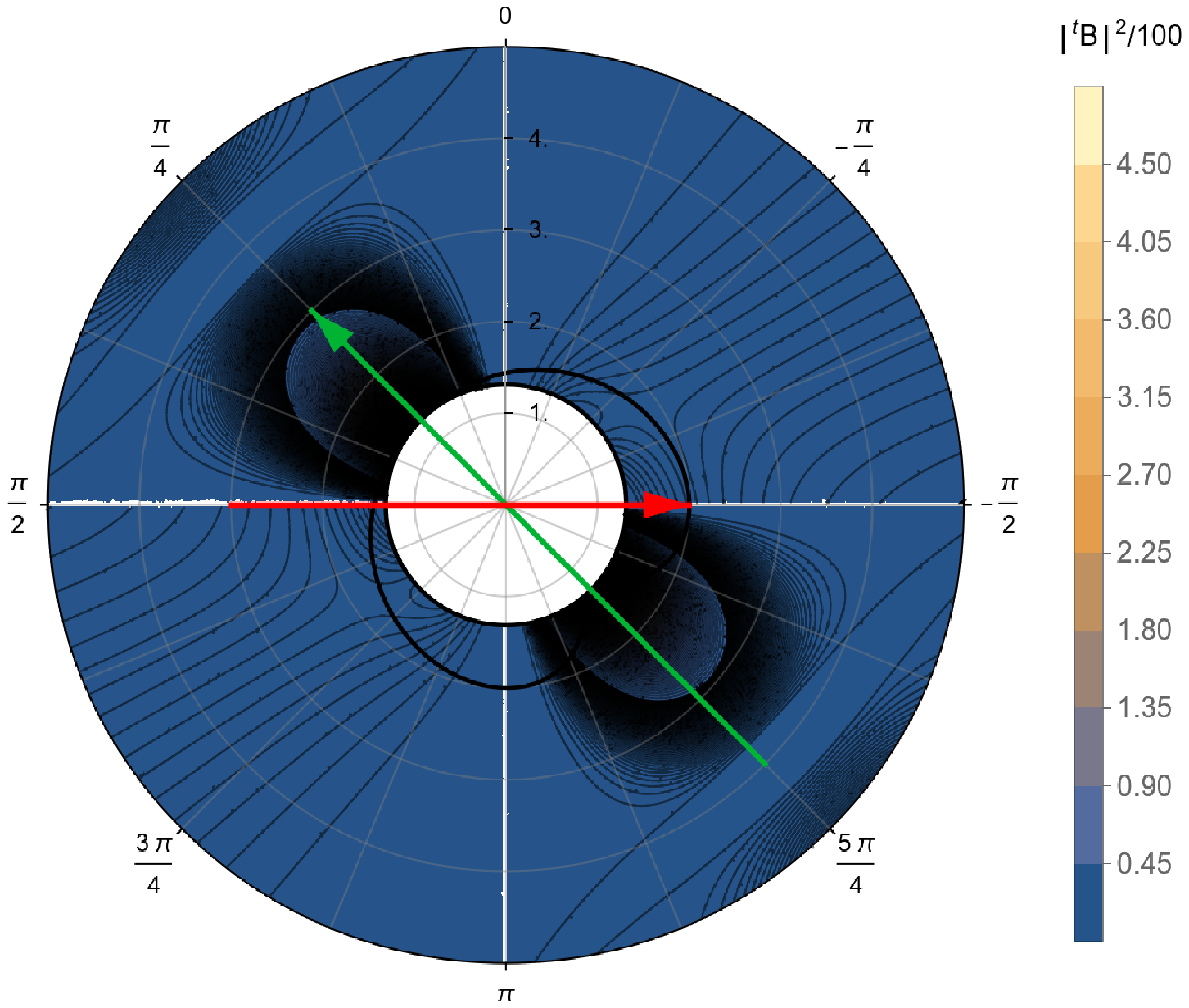}
\includegraphics[height=5cm]{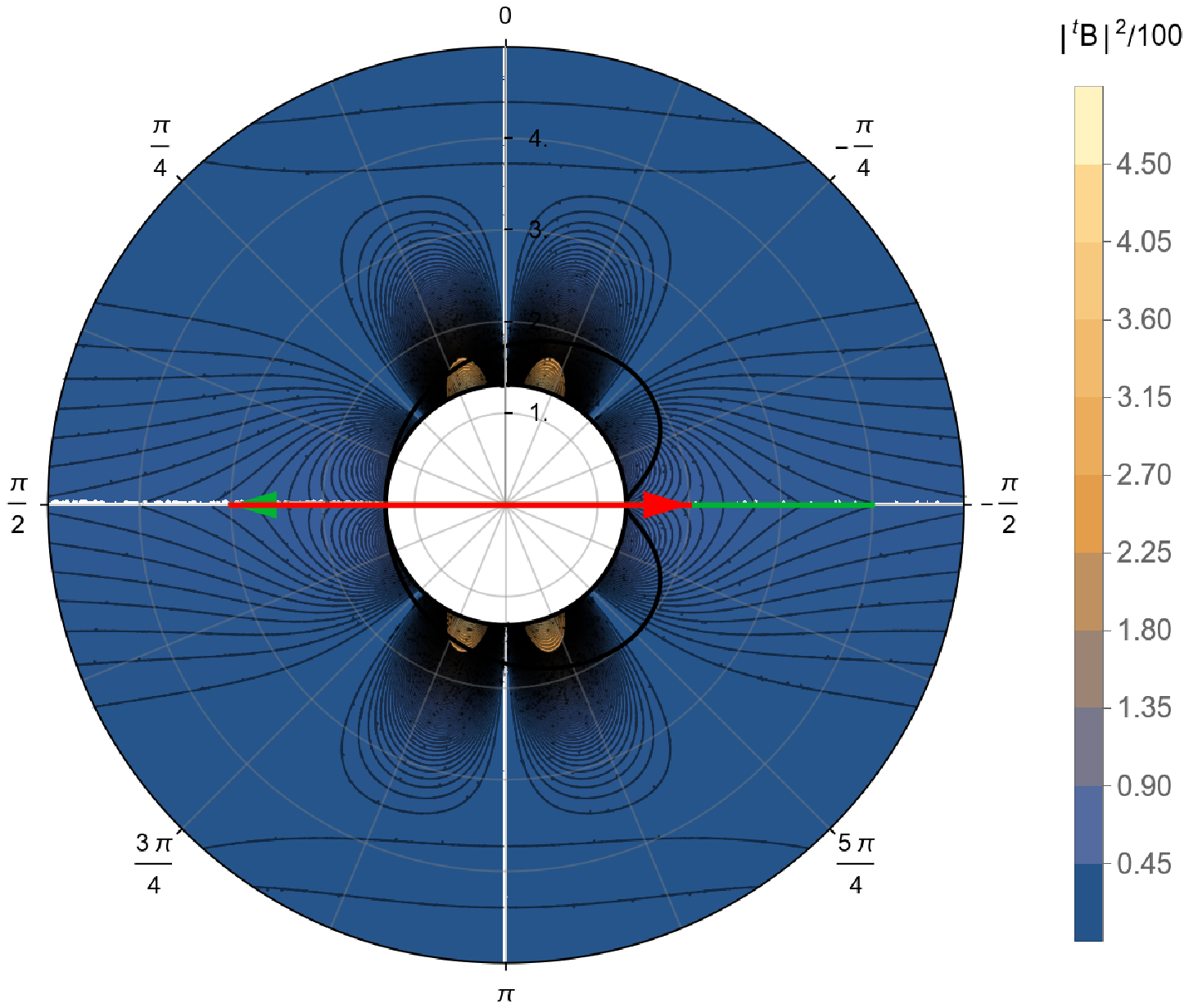}
\caption{Electromagnetic configurations for a Kerr-Schild boosted black-hole for different directions of the boost $\boldsymbol{\hat{n}}$, here represented by the green arrow. The red arrow indicates the direction of the rotation parameter $\omega$. Here $m_0=1, \omega_0=0.95, \gamma=0.900, n_2=0.000$. The most intense radial magnetic field is reached when the boost and the spinning axis are aligned, in which case the space-time and the magnetic field are axially symmetric as in the top-Left panel in which $n_1=1.000, n_3=0.000$. As boost becomes unaligned, with an acute angle less than approximately $\pi/4$. The magnetic field remains intense in the region between the boost and the spinning direction. Exist, near to the orthogonal plane to the boost, an intense magnetic field as shown in the top-Middle panel in which $n_1=0.707, n_2=0.707$. We observe an abrupt decrease in the intensity of the magnetic field when the boost and the rotation axes are orthogonal. The magnetic field iso-lines twist and the magnetic field is, a little bit intense around the rotation axis, as seen in the top-Right panel, for which $n_1=0.000, n_3=1.000$. The magnetic field is weaker with respect to the other configurations when the angle between the boost and the rotation axis is approximately $3\pi/4$. In this situation, with the highest intensity in the plane normal to the boost direction and the minor intensity along the rotation axis, as is displayed in the bottom-left panel, for which $n_1=-0.707, n_2=0.707$. When the boost and the spinning axes are in opposite directions we observe iso-lines of magnetic field aligned with the rotation axes and that appear four lobes around the normal direction to the boost for which the magnetic field is more intense as is shown in the bottom-middle panel, for which $n_1=-1.000, n_2=0.000$.
} 
\label{FigKSvarBoostDir}
\end{figure*}
It is worth mentioning that the intensity of the magnetic field for this configuration has an analogous form to the accretion disk and emission jet measured recently and reported in \cite{science}. We highlight at this point two things about this configuration. Firstly both the two cones axis and the disk orthogonal axis are misaligned. Secondly, the axes of the cones are not directed along the rotation axis.
In Fig. \ref{FigKSvarGamma} we plot the square of the intensity of the spatial part of the purely radial magnetic field $|{^{t}}{\cal B}^r|^2={^{t}}{\cal B}^r \, {^{t}}{\cal B}_r$ for a Kerr-Schild boosted black-hole for a given boost direction, represented by a green arrow, as viewed by an observer with four-velocity $u^{\mu}=(1, 0, 0, 0)$. The section plotted corresponds to the plane containing both the boost and the spinning vectors. The spin direction for the black hole is represented by a red arrow. We observe that magnetic field intensity along the boost direction increases as the boost $\gamma$ becomes higher. Also, along the spinning axis a shear-type effect on the intensity iso-lines of the magnetic field becomes stronger as boost parameter $\gamma$ increases. The most intense region in the plots are located between the boost and the spinning vectors, which form a conical structure. However, it is worth noticing an emission in the normal direction to the boost vector, corresponding to a section of a disk with is approximately orthogonal to the boost vector.      
In Fig. \ref{FigKSvarBoostDir} we show the square of the intensity magnetic field $|{^{t}}{\cal B}^r|^2$ for different orientations of the boost vector. We observe that when the boost and the rotation axis are aligned, the magnetic field becomes symmetrical, being more intense around the rotation axis and less intense in the plane orthogonal to the rotation axis. This is the configuration in which the magnetic field intensity is maximum. The magnetic field intensity decreases as a whole for other boost directions as the angle between the boost and the spinning increases. The magnetic field turns symmetric again when the boost and the rotation directions becomes anti-parallel, in which case four lobes of intense magnetic field appear around the equator (the plane normal to the rotation axis). In addition, low intensity iso-lines aligned with the direction of the rotation axis appear. It is important to highlight that in any case the magnetic field cease to be radial.   
In Fig. \ref{FigKSvarBoostInten} we mantain fixed the mass $m_0=1.0$, the boost parameter $\gamma=0.9$, the boost and the rotation directions $\boldsymbol{\hat{n}}$ and $\boldsymbol{\hat{z}}$ respectively. We explore the intensity of the magnetic field as the rotation parameter $\omega_0$ vary. We note that for a low rotation parameter $\omega_0=0.005$, the magnetic field is weak $|{^{t}}{\cal B}^r|^2\le 5\times 10^3$ (See the Top-Left panel). As $\omega_0$ increases, the magnetic field becomes stronger, unitl $\omega_0\simeq 0.6$ (See the Top-Right panel). After this limit, the intensity of the magnetic field decreases for the range plotted $r=6$. A bubble region of zero magnetic field in the proximity of the ergosphere around the boost direction becomes bigger as $\omega_0$ aproximates to $\omega_0=1$.    
\begin{figure*}%[h!]
\centering
\includegraphics[width=5cm,height=5cm]{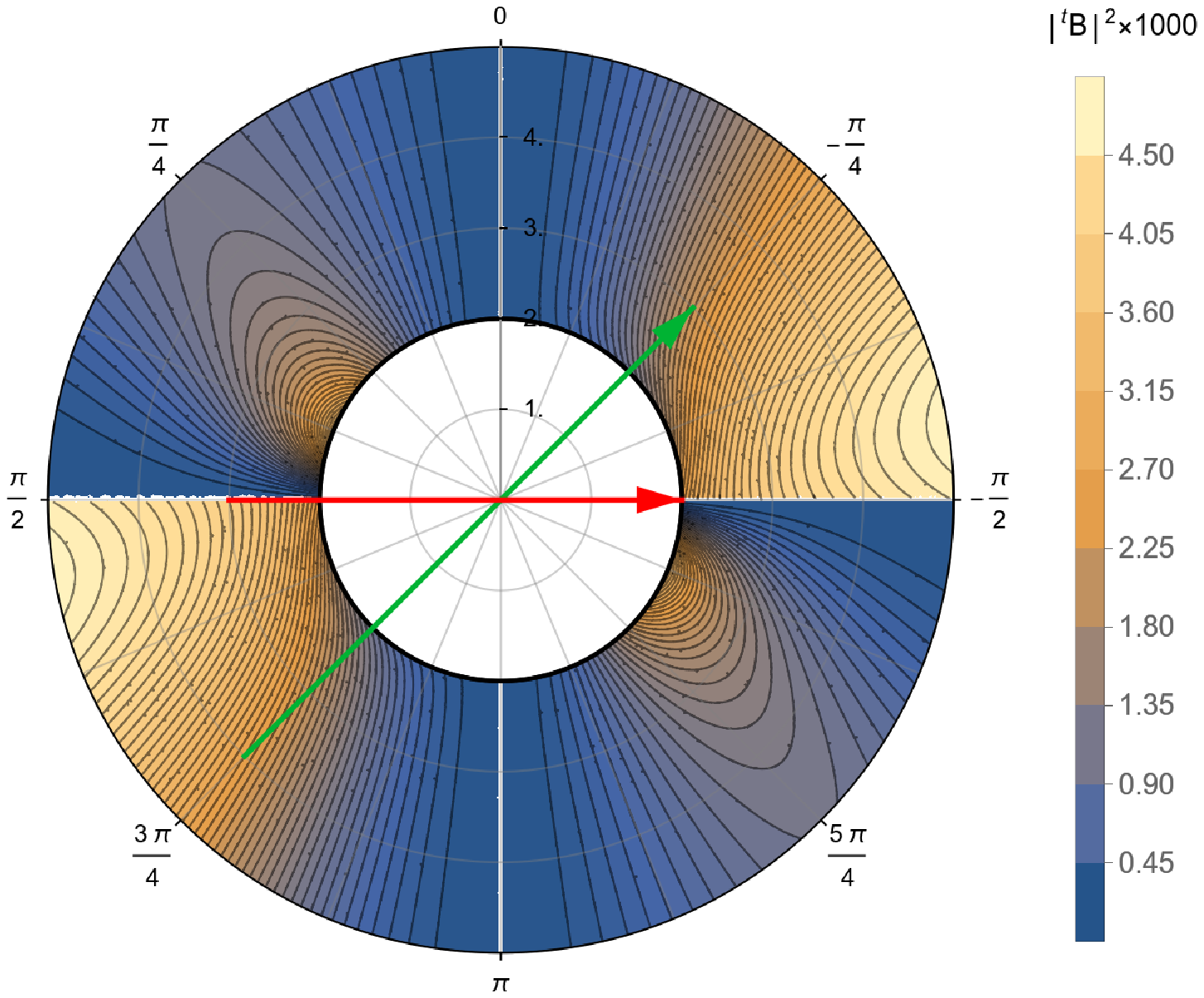}
\includegraphics[width=5cm,height=5cm]{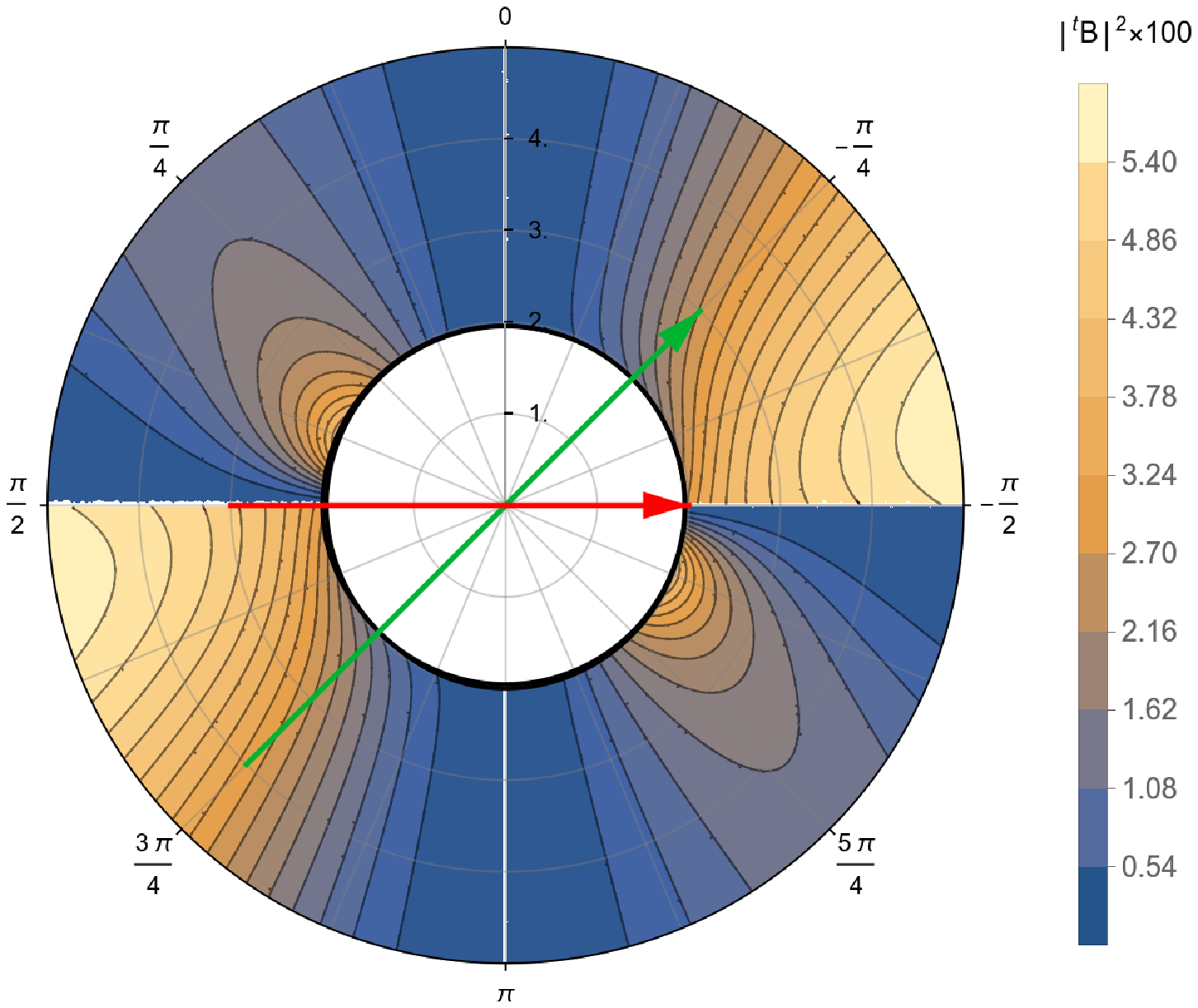}
\includegraphics[width=5cm,height=5cm]{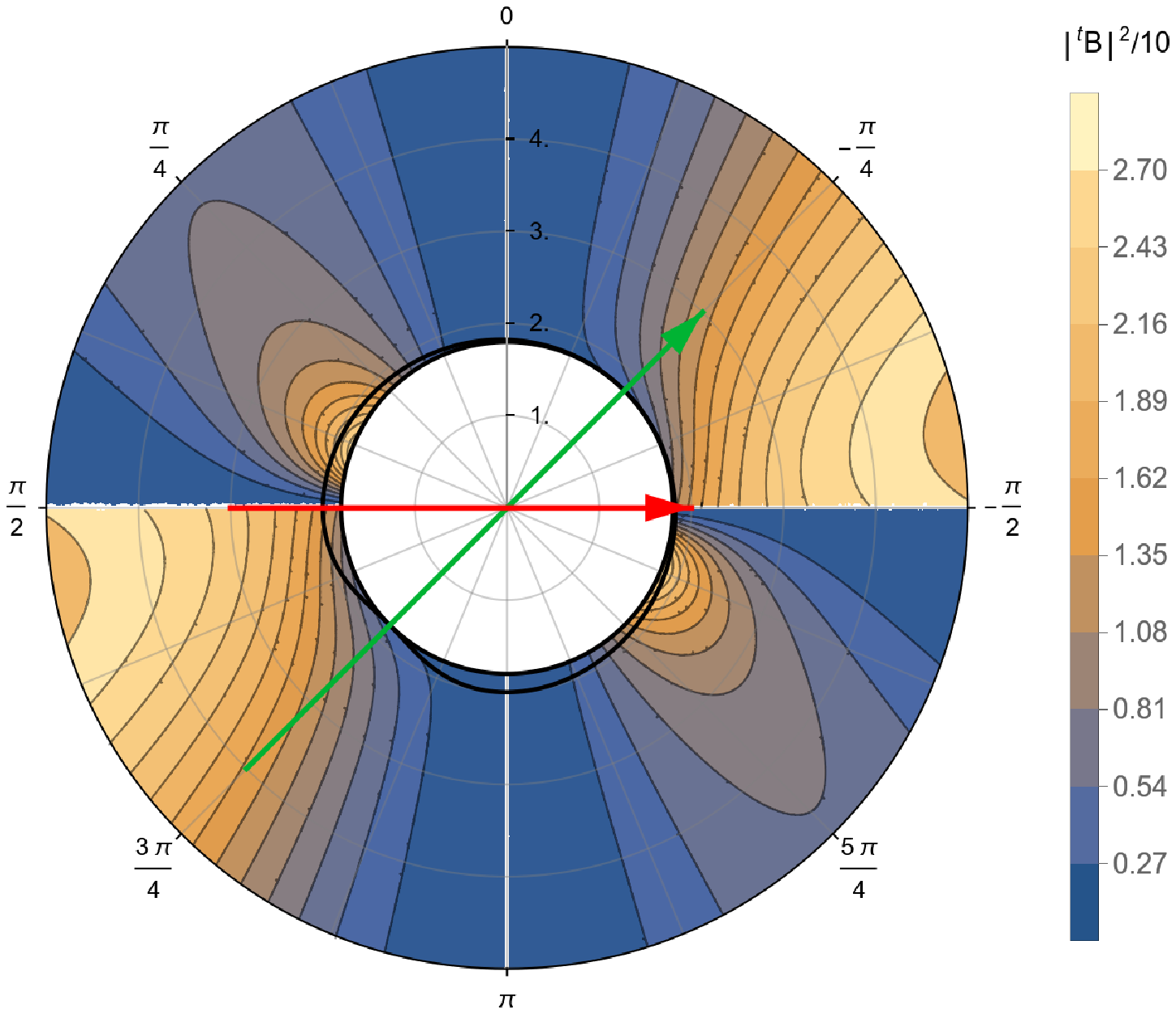}\\
\includegraphics[width=5cm,height=5cm]{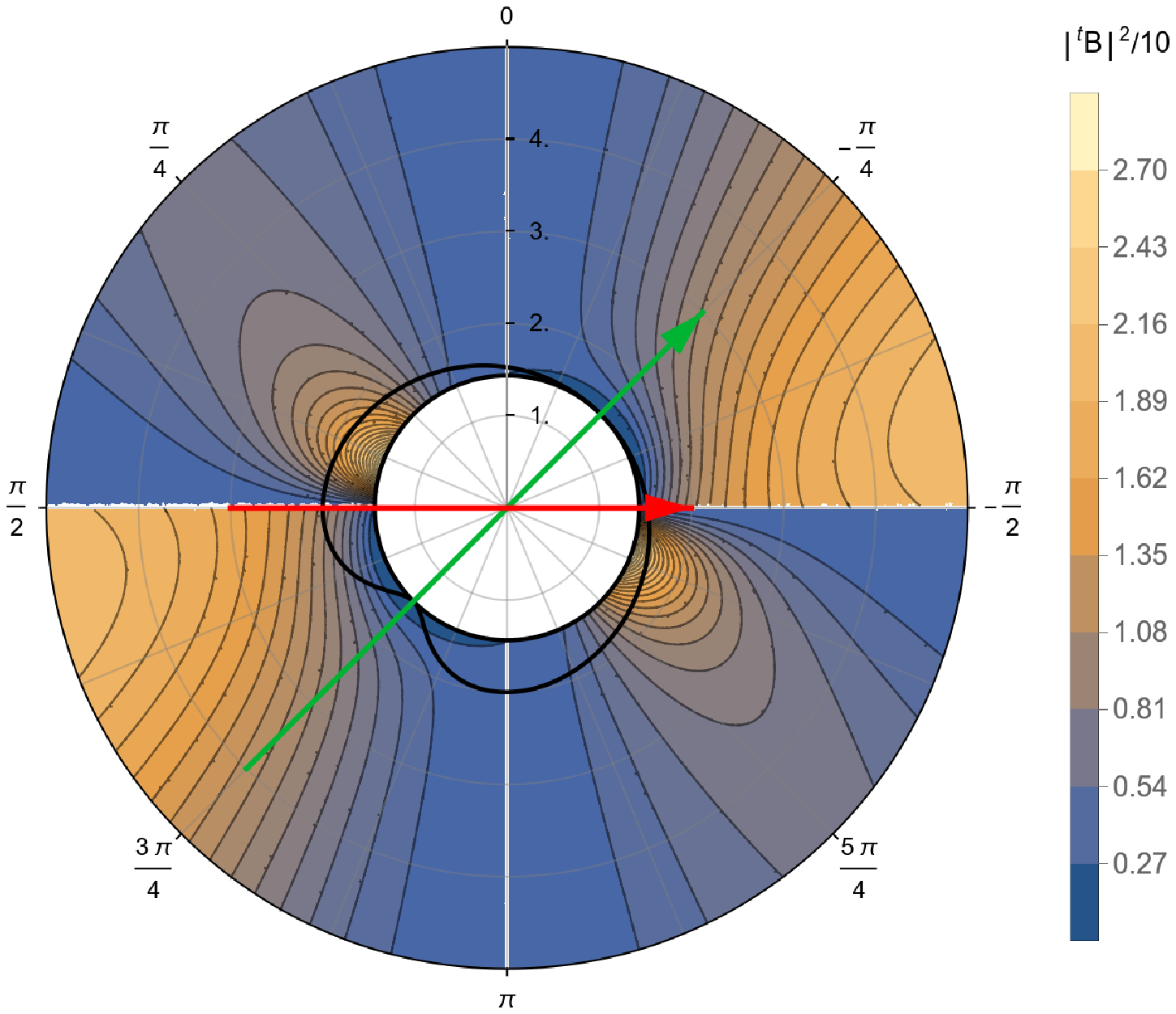}
\includegraphics[width=5cm,height=5cm]{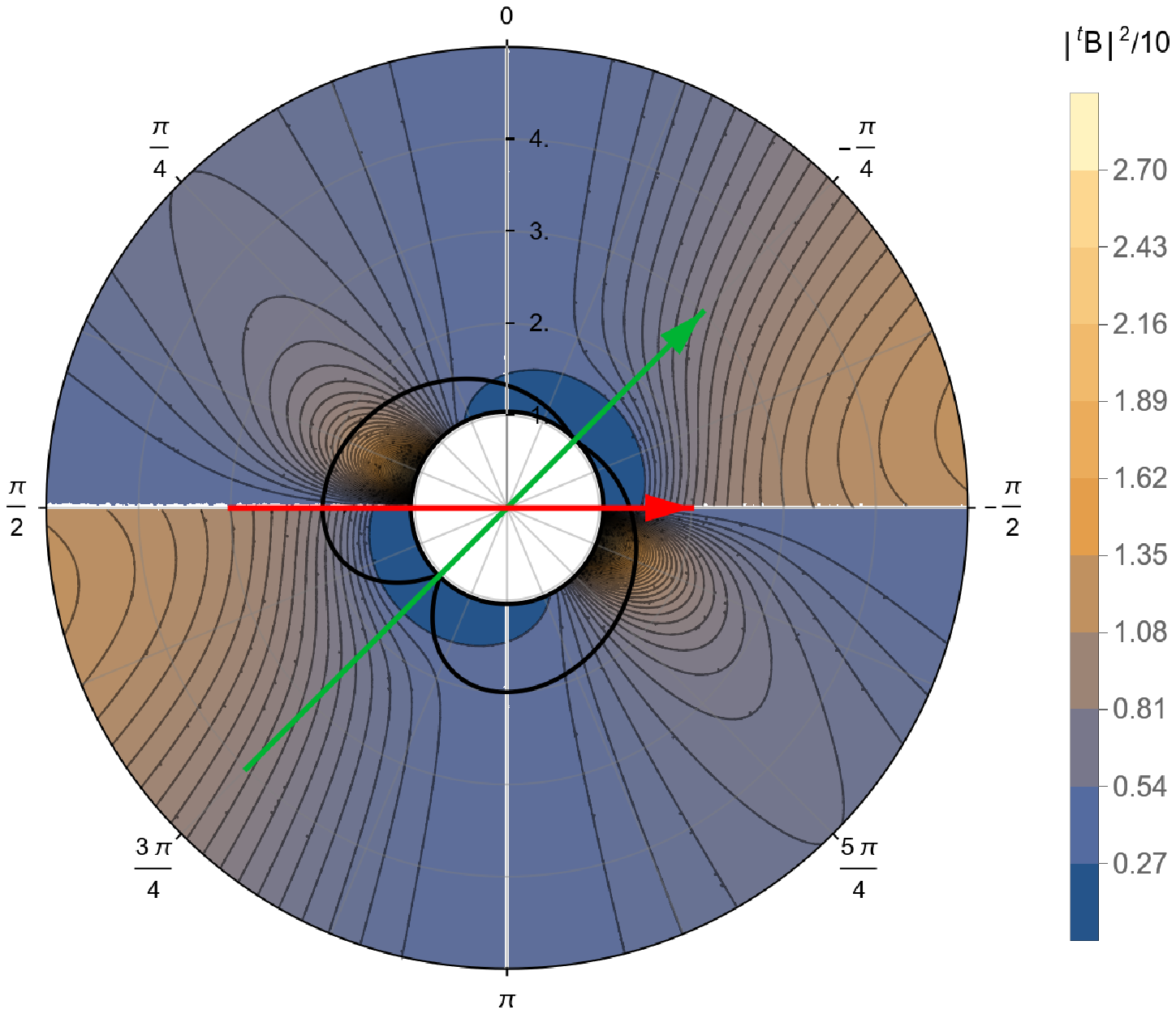}
\caption{Density plots for the square of the intensity of magnetic field $|{^{t}}{\cal B}^r|^2$ for a Kerr-Schild boosted black-hole for a given direction of the boost $\boldsymbol{\hat{n}}$ and for the boost parameter $\gamma$ and different values of the rotation parameter $\omega$. The boost and the spinning directions are represented by the green and the red arrows, respectively. Here $m_0=1, \gamma_0=0.90, n1=0.7071, n_2=0.000, n3=0.7071$. For a very low rotation parameter $\omega_0=0.05$ as in the top-left panel we observe a magnetic field of very low intensity. A rapid increase of the magnetic field intensity is given as the rotation parameter grows. We observe that the intensity is multiplied by $10$ in the top-middle panel, in which $\omega_0=0.3$, and by $100$ in the top-right panel for which $\omega_0=0.6$. As the spinning increases, the ergosphere becomes bigger, the horizon shrinks, and the magnetic field decreases as shown in the Bottom-left panel for which $\omega_0=0.9$. For higher values of the rotation parameter, $\omega_0=0.999$, a zero magnetic field region emerges around the boost direction as shown in the Bottom-middle panel. 
} 
\label{FigKSvarBoostInten}
\end{figure*}

Applying the transformations (\ref{rit1n}) together with $u=t-r$, the components of the $4$-vector ${^{t}}{\cal B}^\mu$ measured in BS coordinates read
\begin{eqnarray}
^U{\cal B}^U&\simeq& K(\theta, \phi)({^{t}}{\cal B}^t-{^{t}}{\cal B}^r)\\
^U{\cal B}^R&\simeq& \frac{1}{K(\theta, \phi)}{^{t}}{\cal B}^r.
\end{eqnarray}
We compute the norm of the spatial part for this magnetic field as $|\,^U{\cal B}^R|^2= \,^U{\cal B}_R \,^U{\cal B}^R$. In Fig. \ref{FigBSvarGamma0} we vary the boost parameter $\gamma$, fixing the mass, the boost direction and the rotation parameter as $m_0=1, n_1=0.7071, n_2=0.000, n_3=0.7071$. The boost and the spin directions are represented by the green and the red arrows respectively. We note that the intensity of the magnetic field increases as the boost parameter grows. For this configuration of the boost direction, the magnetic field is concentrated between the boost and the spin vectors. However, magnetic field appears as emanated in the orthogonal direction to the boost. This behaviour is intensified as the boost increases. A region of zero magnetic field emerges in the proximity of the ergosphere, in the boost direction. A shear-like effect along the rotation axis on the intensity iso-lines is intensified as the boost parameter increases. However, the spatial component of the magnetic field on Bondi-Sachs is purely radial. 
\begin{figure*}%[h!]
\centering
\includegraphics[width=5cm,height=5cm]{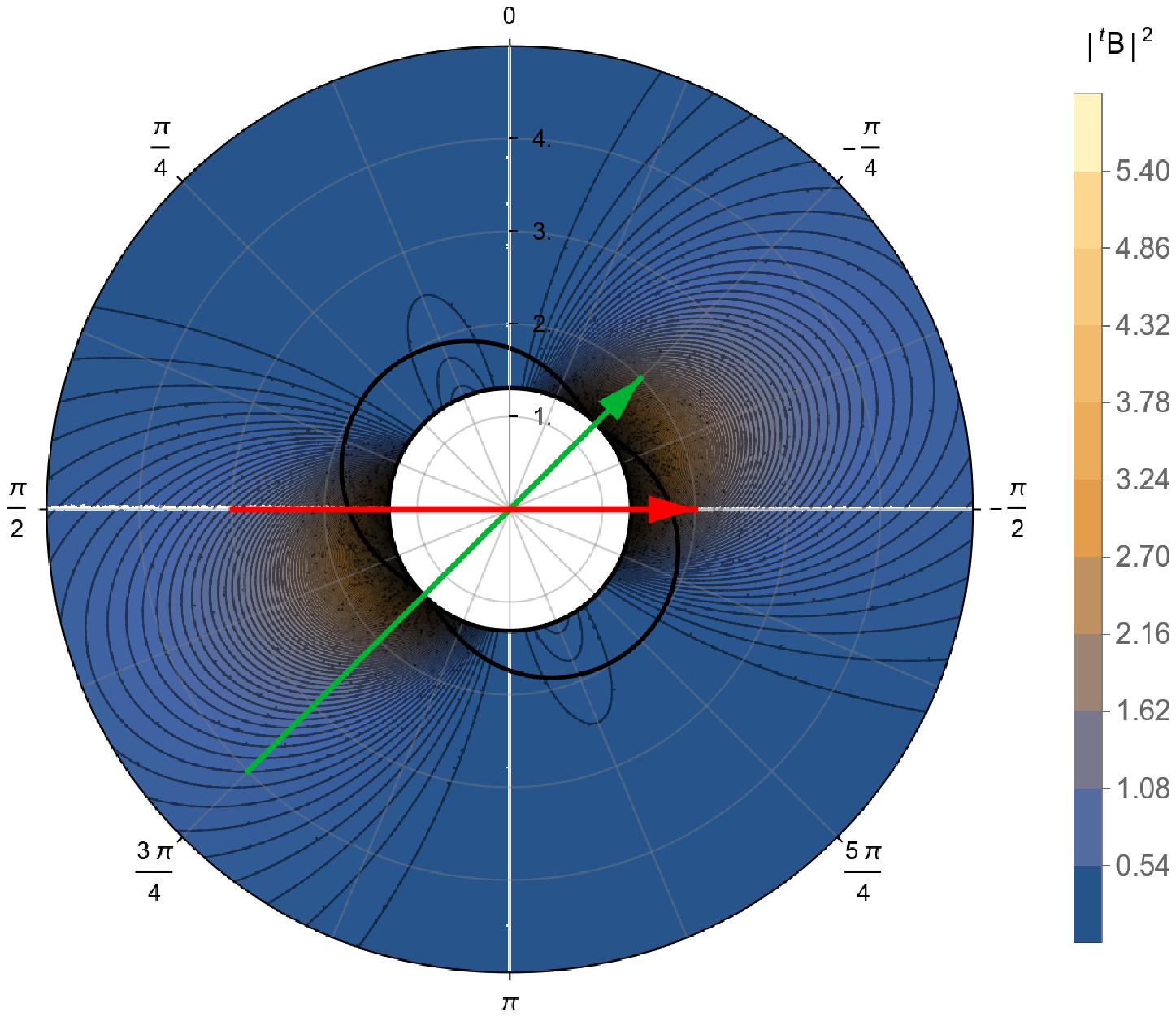}
\includegraphics[width=5cm,height=5cm]{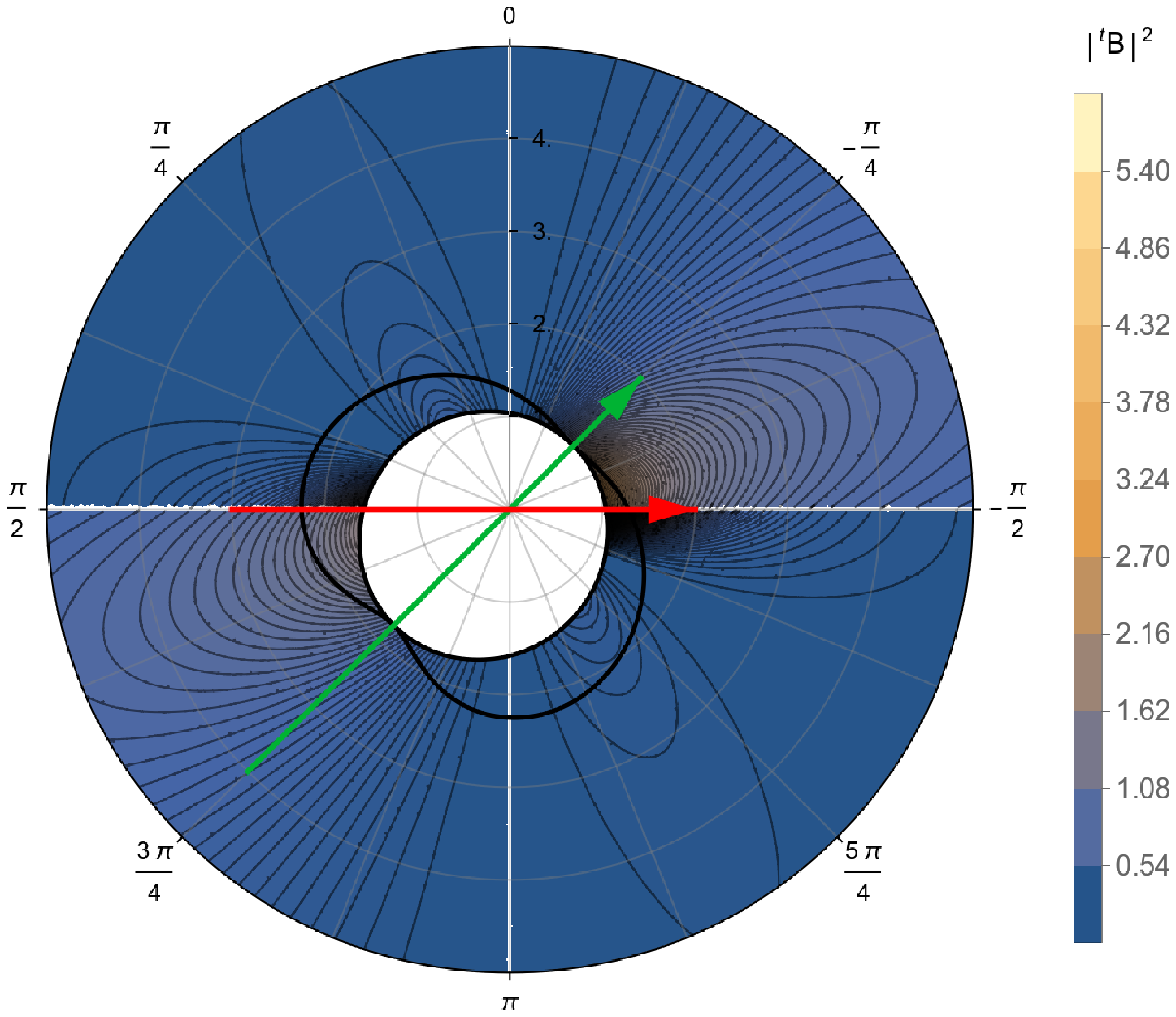}
\includegraphics[width=5cm,height=5cm]{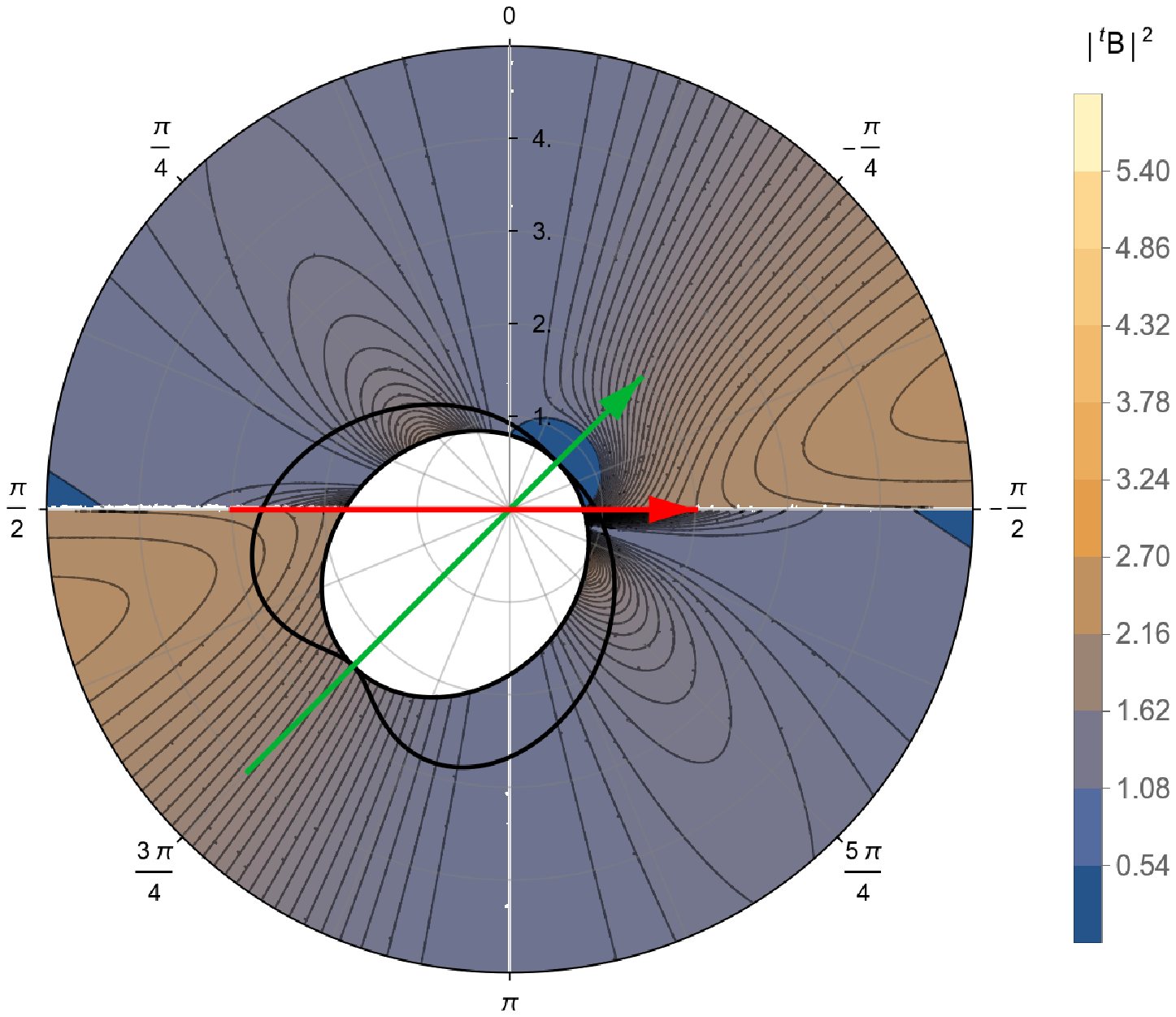}
\caption{Configurations of the magnetic field for a Bondi-Sachs boosted black-hole for different values of $\gamma$ as seen by an observer with four-velocity in Kerr-Schild coordinates $u=(1,0,0,0)$. Here the mass, the rotation parameter and the boost direction are fixed, $m_0=1, \omega_0=0.95, n1=0.7071, n_2=0.000, n3=0.7071$. We plot $|{^{U}}{\cal B}^r|^2$ and observe a shear-like effect on the iso-lines of the magnetic field, just as in the Kerr-Schild case. For no boost, $\gamma=0.00$, there is no shear effect as seen in the left panel. As the boost parameter increases, $\gamma=0.30$, the intensity of the magnetic field and the shear become more evident, as displayed in the middle panel. For a boost, $\gamma=0.60$, we observe regions of zero magnetic field. One of these regions is close to the horizon, around the boost direction. The other region appears close to the rotation axis, far away from the ergosphere. The magnetic field is intense in the region between the boost and the rotation axis but appears an intense magnetic field region close to the plane orthogonal to the boost direction, as shown in the right panel.
} 
\label{FigBSvarGamma0}
\end{figure*} 
In Fig \ref{FigBSvarw0} we plot $|{^{U}}{\cal B}^r|^2$ for a Bondi-Sachs black hole for different values of the spinning parameter $\omega_0$, conserving as constants the mass, the boost parameter and the boost direction, fixed as $m_0=1, \gamma_0=0.900, n1=0.7071, n_2=0.000, n3=0.7071$ respectively. The intensity of the magnetic field becomes stronger as $\omega_0$ increases. For the configuration showed in this figure, the region with zero magnetic field, in the proximity to the ergosphere, remains unaltered. Also, it is worth to note that as $\omega_0$ grows, the magnetic field in the region between the boost and the rotation vectors intensifies with respect the magnetic field in other parts of the plotted region. The same occurs with the magnetic field emanated in the orthogonal direction to the boost direction.    
\begin{figure*}%[h!]
\centering
\includegraphics[height=5cm]{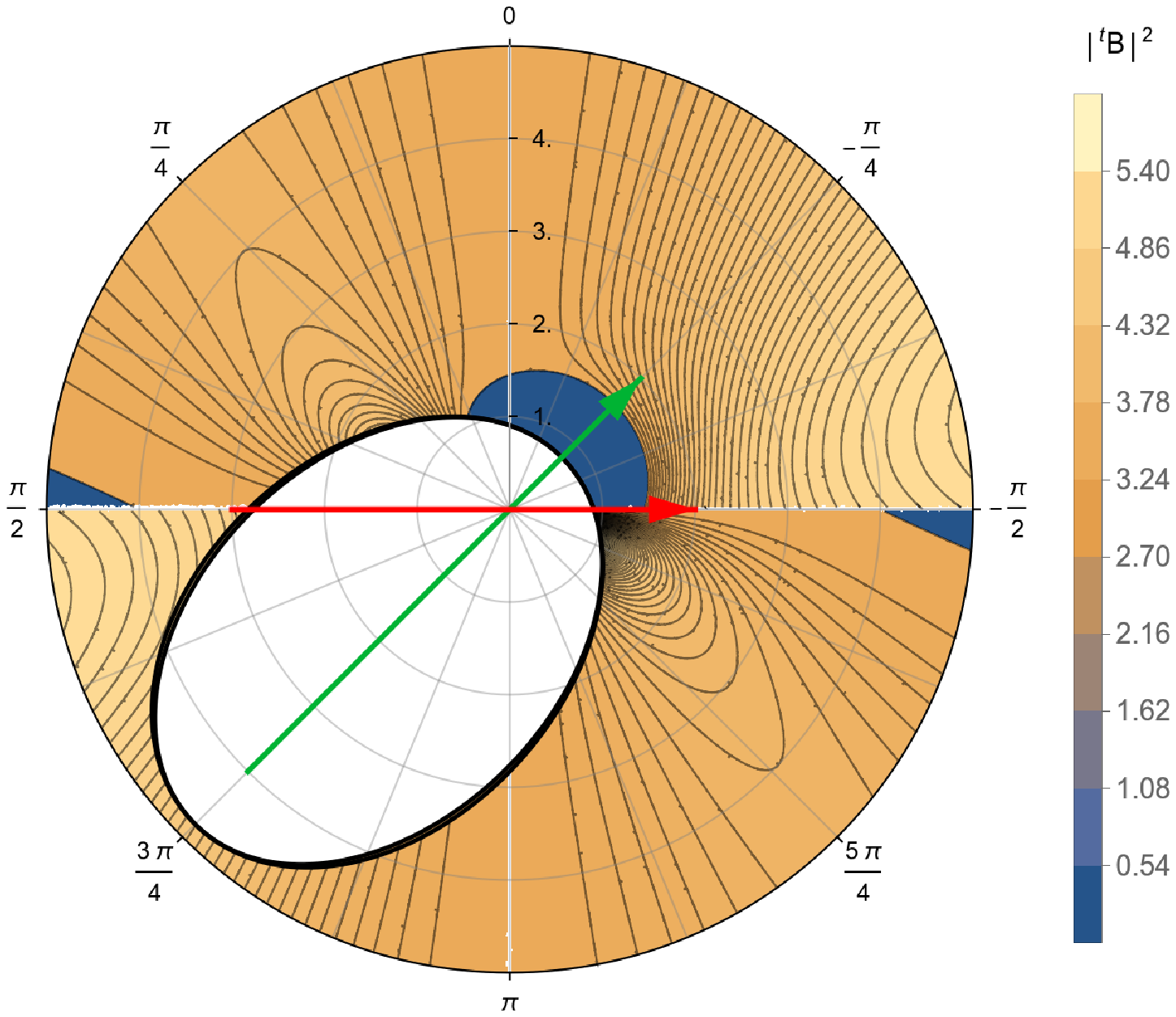}
\includegraphics[height=5cm]{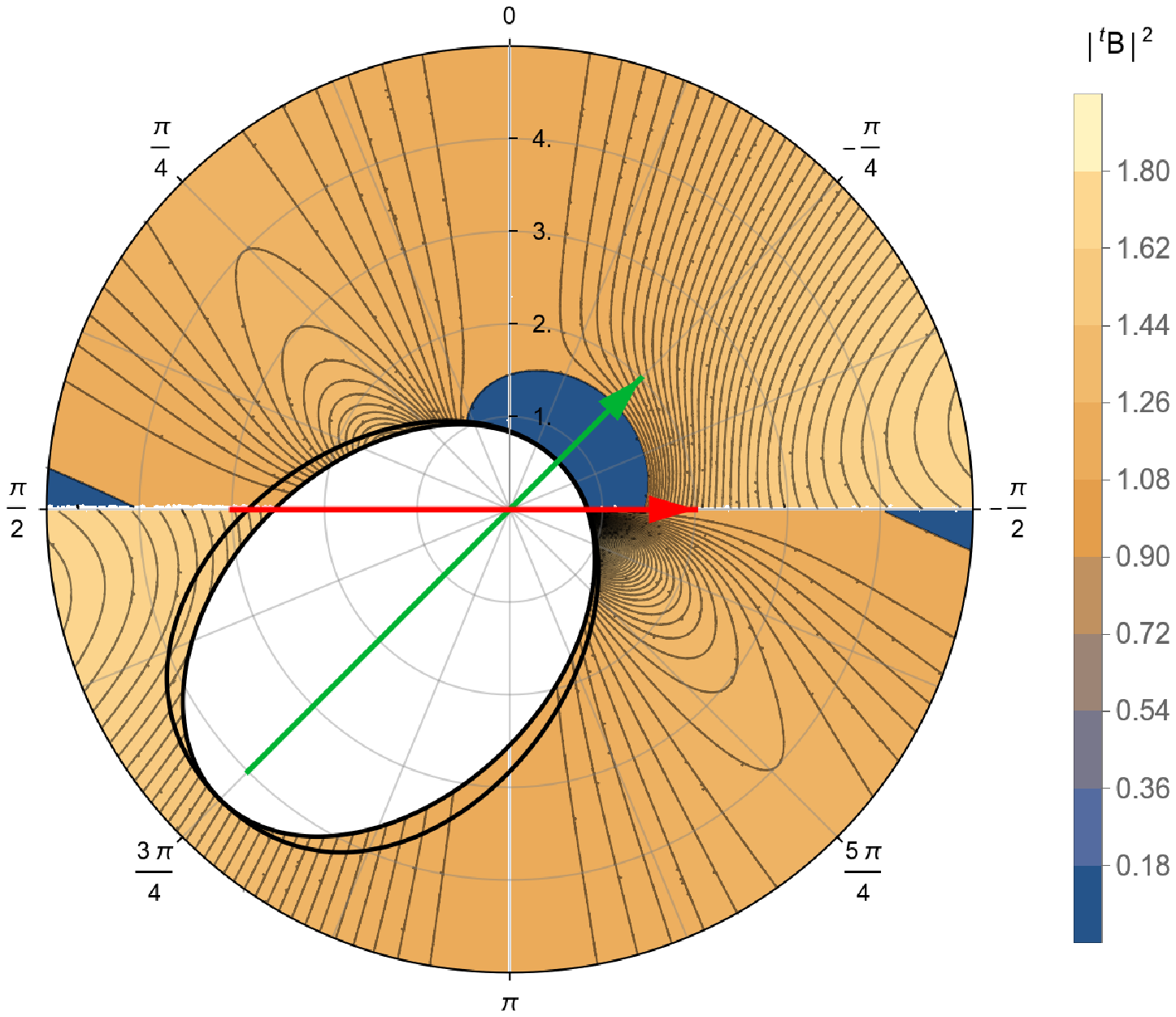}
\includegraphics[height=5cm]{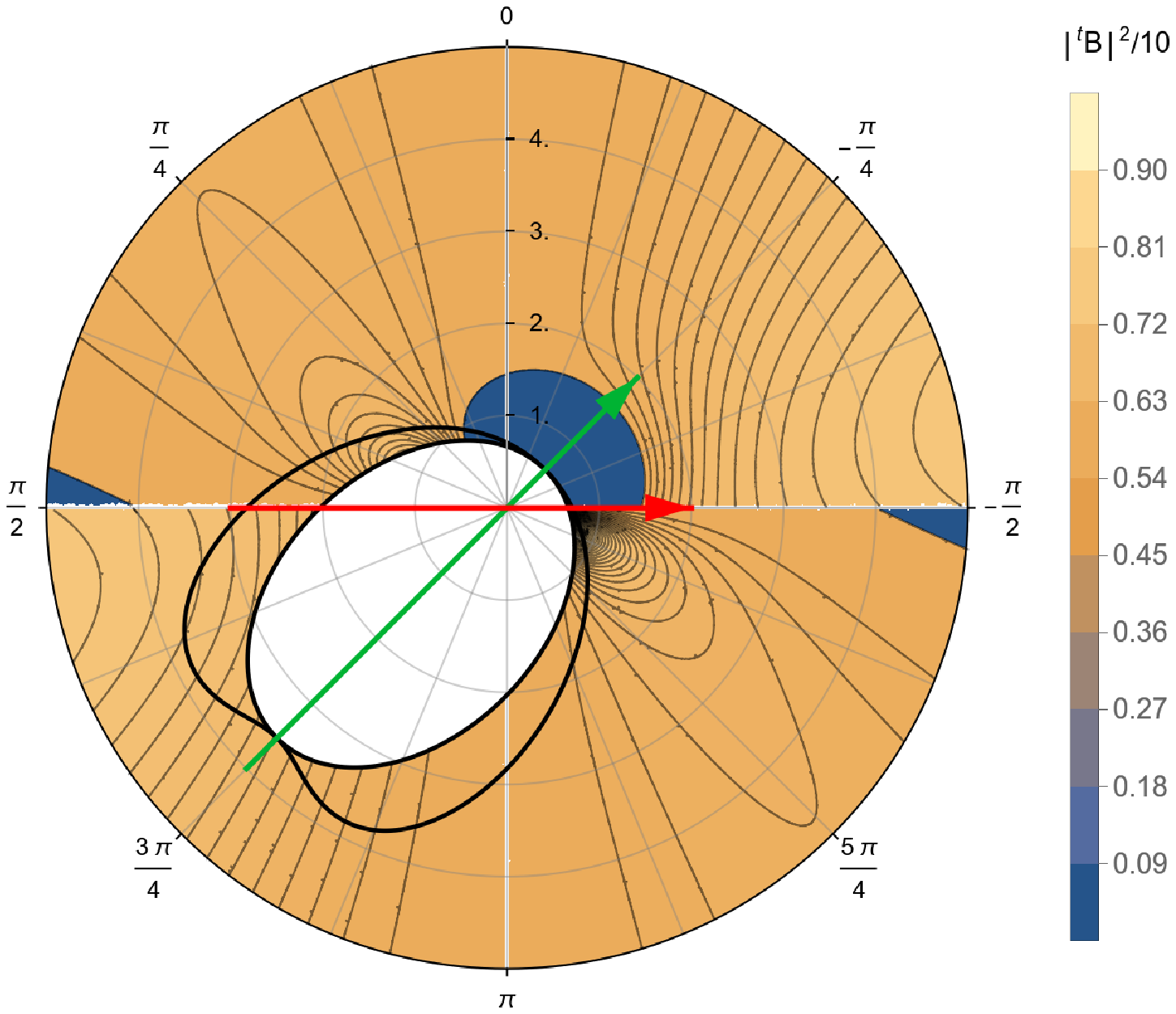}
\caption{Density plots for the square of the intensity of magnetic field $|{^{U}}{\cal B}^r|^2$ for a Bondi-Sachs boosted black-hole for different values of the spinning parameter $\omega_0$. Here the mass, the boost parameter and the boost direction are fixed $m_0=1, \gamma_0=0.900, n1=0.7071, n_2=0.000, n3=0.7071$. We observe that the intensity of the magnetic field decreases as the rotation parameter $\omega$ increases. The region of zero magnetic field close to the horizon, around the boost direction, remains unaltered, however the two regions of zero magnetic field around the spiining axis tends to dissapear as the rotation parameter grows. Left panel: $\omega_0=0.30$, middle panel: $\omega_0=0.60$ and right panel: $\omega_0=0.90$.
} 
\label{FigBSvarw0}
\end{figure*} 

In Fig. \ref{FigBSvarn} we observe that the intensity of the magnetic field is maximum when the boost and the spin axis formed an acude angle of approximately $\pi/4$, and decreases as the angle between the boost and the rotation axis increases. In the configuration in which the boost and the spin axis are aligned a region without magnetic field appears around the direction in which the black hole is boosted. For the configuration for which the boost is anti-parallel to the spin axis we observe that five lobes appear near to the boost direction. In this configuration, far from the black-hole ergosphere, despite weak and radial, the magnetic field remains with the same intensity along the rotation axis and the magnetic iso-lines become parallel to the rotaion axis.  
\begin{figure*}%[h!]
\centering
\includegraphics[height=5cm]{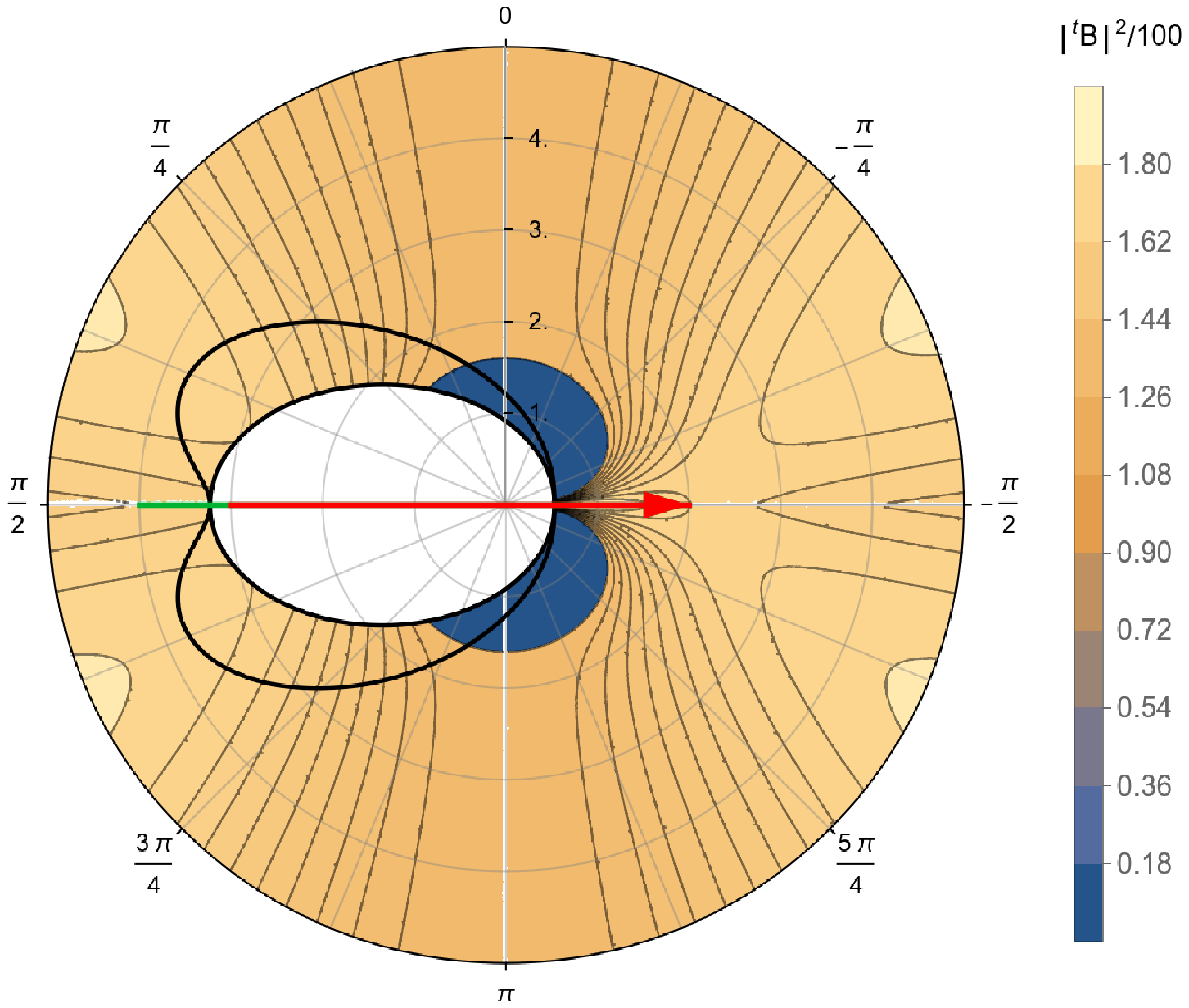}
\includegraphics[height=5cm]{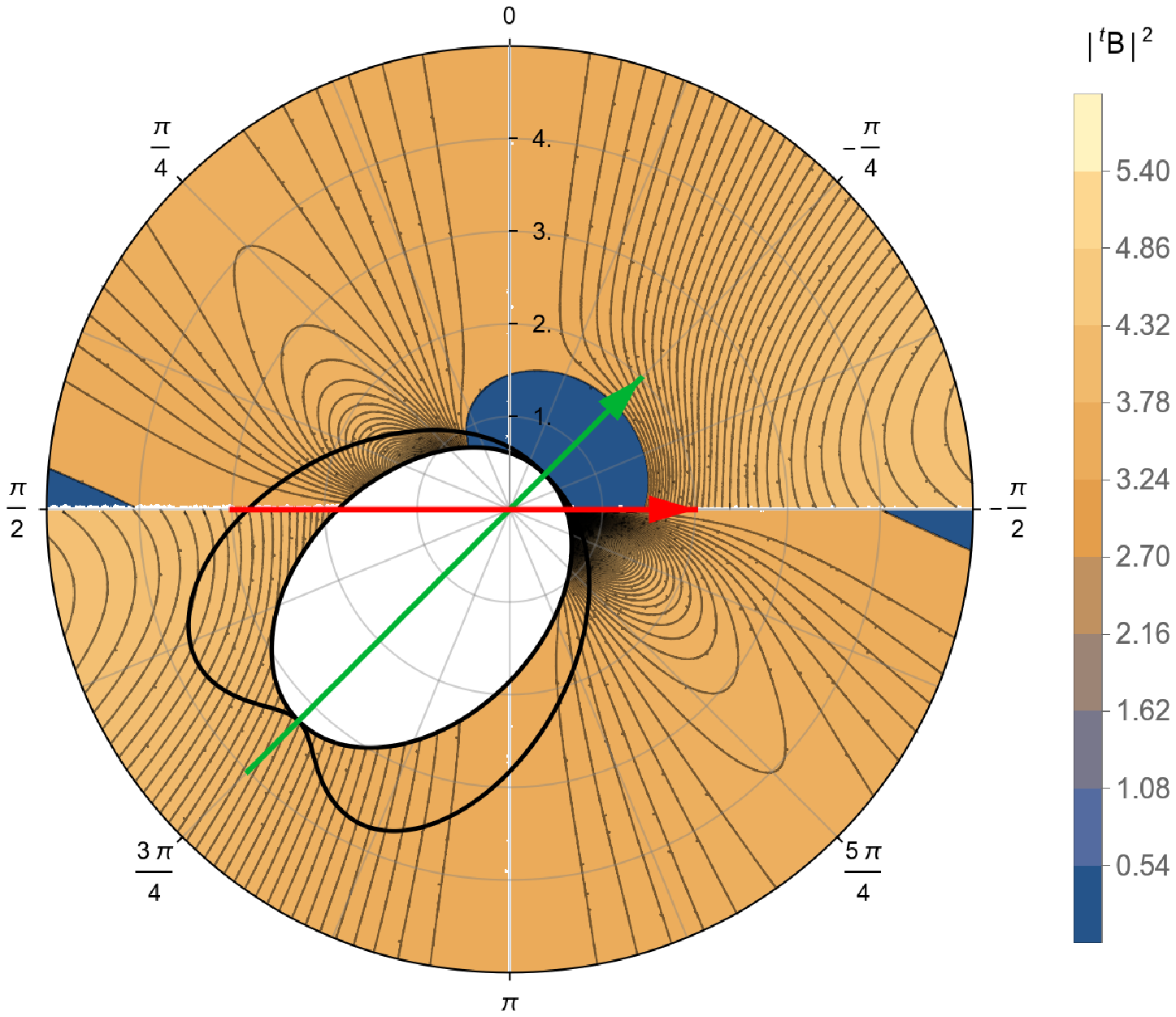}
\includegraphics[height=5cm]{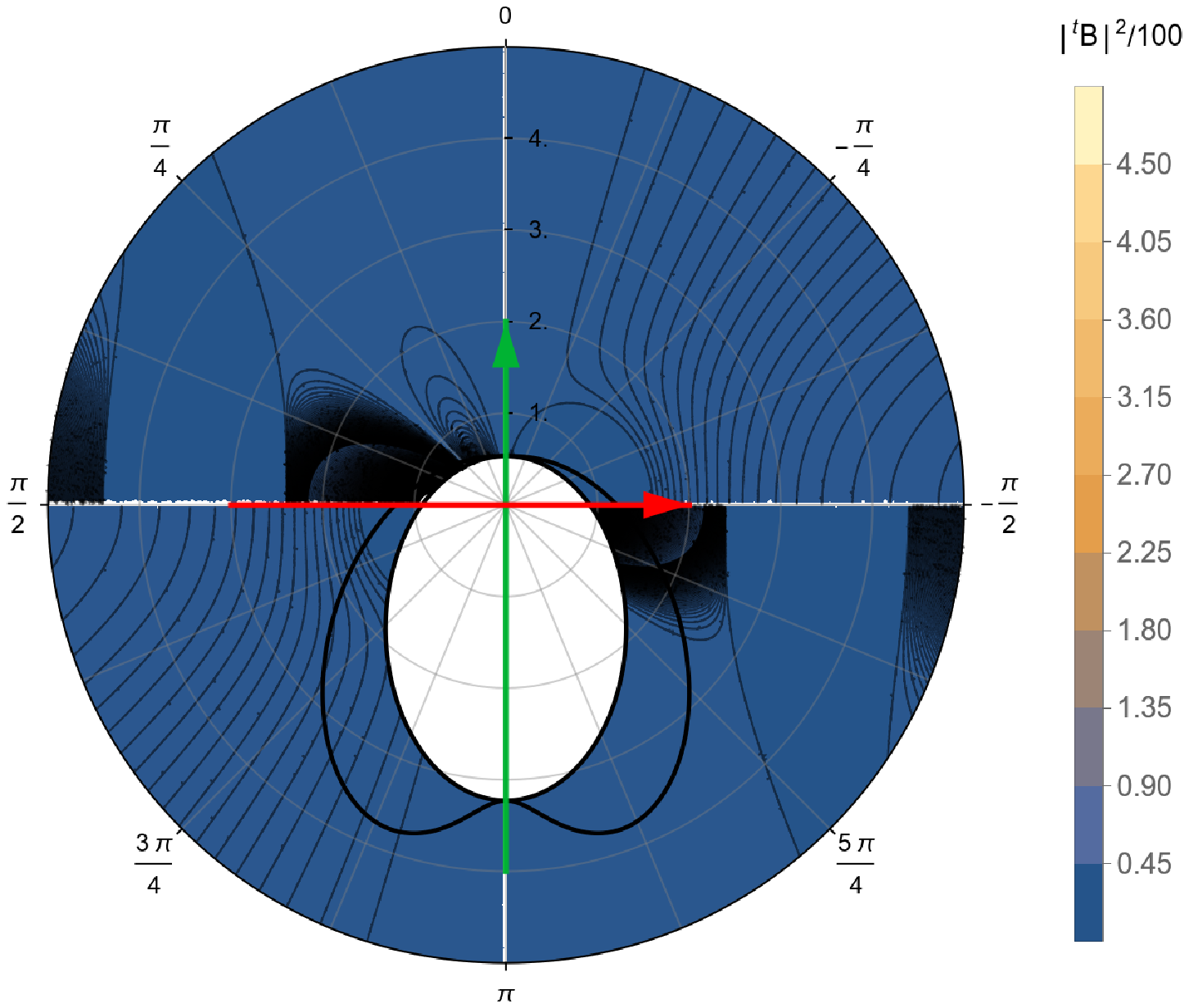}\\
\includegraphics[height=5cm]{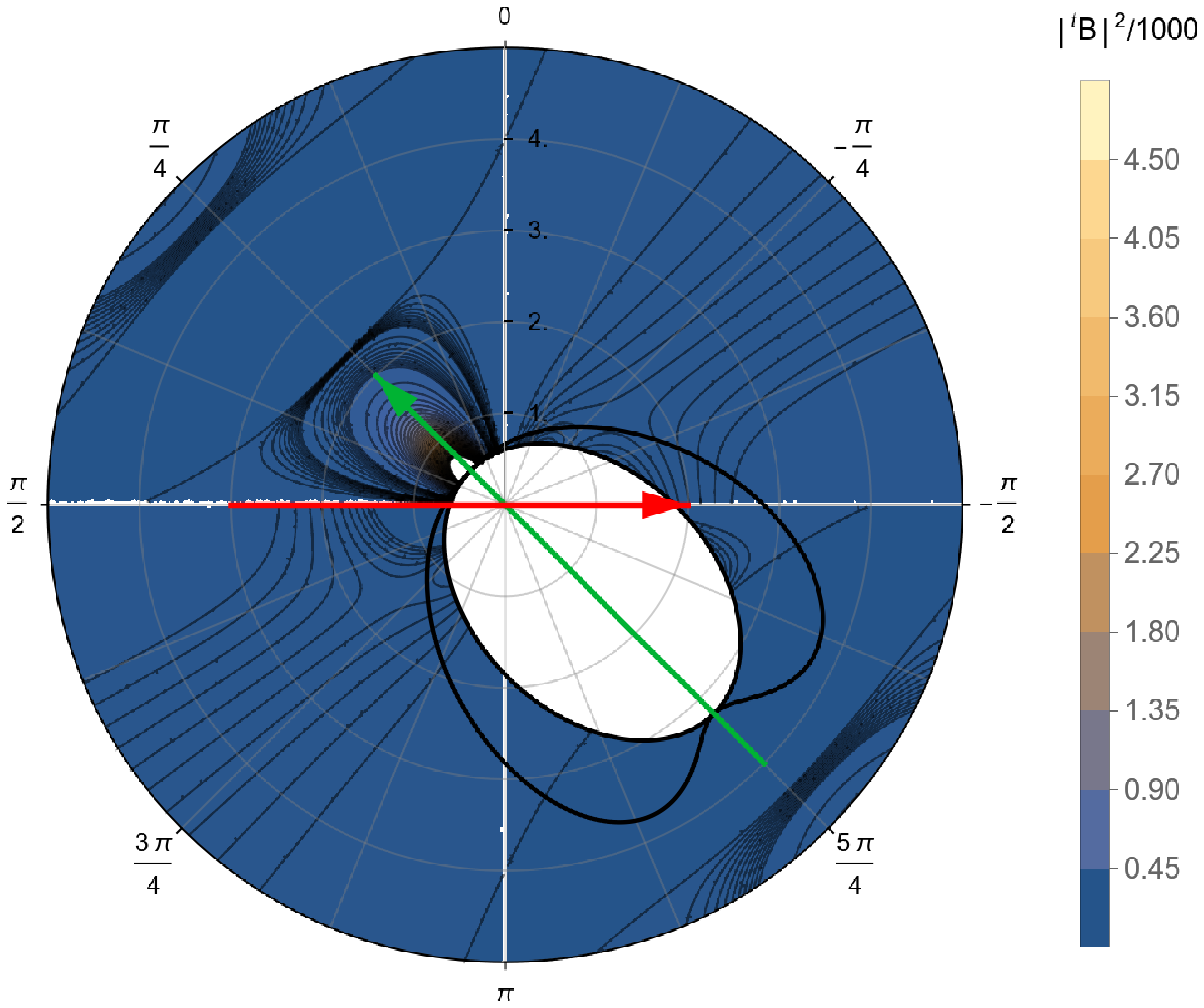}
\includegraphics[height=5cm]{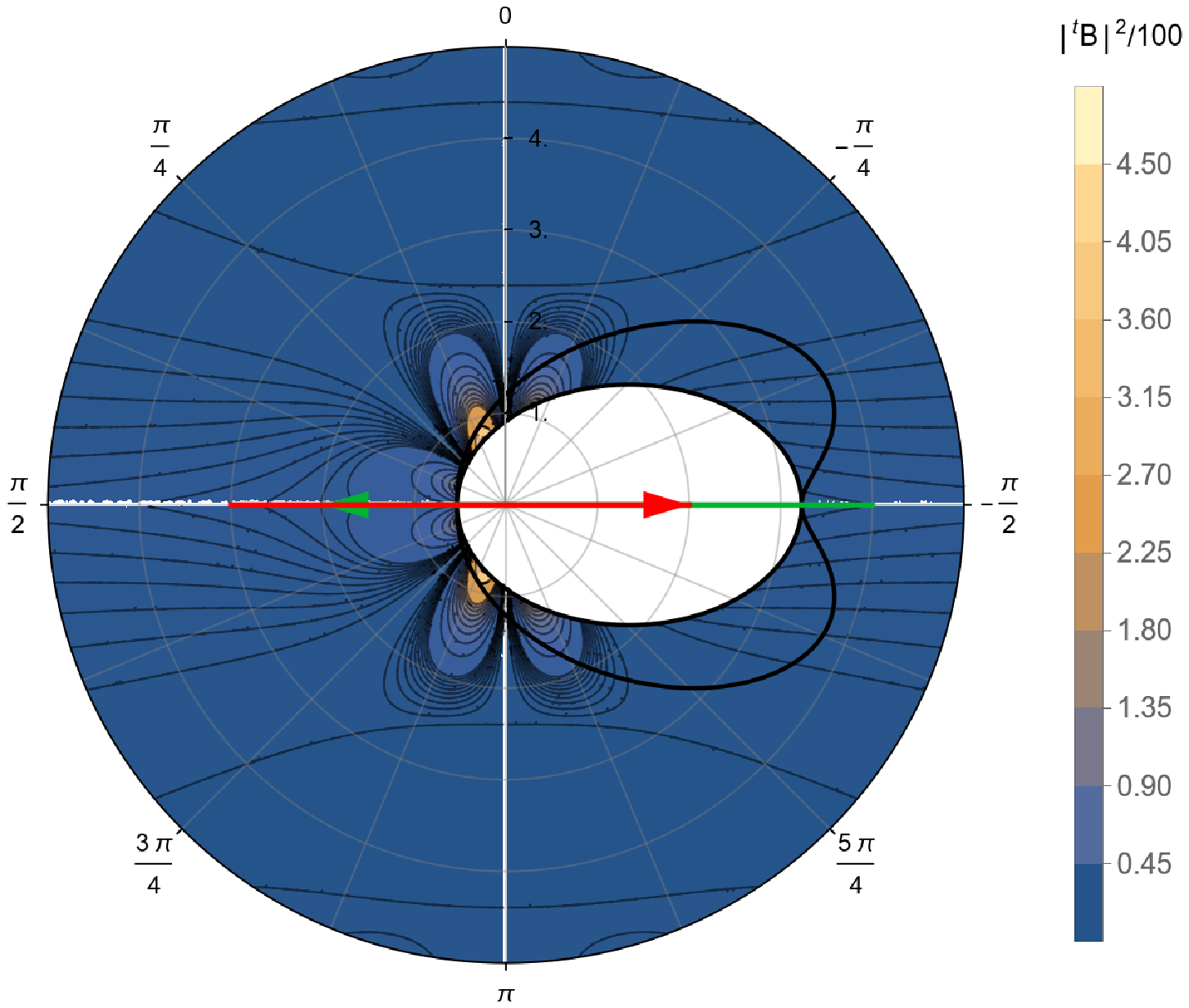}
\caption{Configurations of the purely radial magnetic field for a Bondi-Sachs boosted black-hole for different boost directions $\boldsymbol{\hat{n}}$. Here the mass, the boost and the rotation parameter are constant, with values $m_0=1, \gamma_0=0.900, \omega_0=0.95$. The configuration for axially symmetric magnetic field is when the boost and the spinning axis are aligned, as seen in the top-left panel, in which $n_=1.00, n_2=0.00, n_3=0.00$. The maximum intensity magnetic field is reached when the boost and the rotation axes form an acute angle of approximately $\pi/4$. The magnetic field intensity is higher in the regions between the rotation and the boost axes, however, there is another notorious region of high intensity, close to the plane perpendicular to the boost direction, as shown in the top-middle panel, for which $n_1=0.7071, n_2=0.00, n_3=0.7071$.  The magnetic field decreases abruptly when the boost and the rotation axis are orthogonal, but still, in this situation, the maximum intensity in this configuration is in the region between the boost and the rotation axis as seen in the top-right panel, for which $n_1=0.00, n_2=0.00, n_3=1.00$. The magnetic field with minimum intensity value is reached when the boost and the rotation axes form an angle of approximately $3\pi/4$. In this situation, the magnetic field is slightly more intense in the boost direction, near to the horizon, as seen in the bottom-left panel, for which $n_1=-0.7071, n_2=0.00, n_3=0.7071$. When the boost and the rotation direction are in opposite directions, six lobes around the boost direction are formed, with the second lobes more intense than the other, as displayed in the bottom-right panel, $n_1=-1.00, n_2=0.00, n_3=0.00$.
} 
\label{FigBSvarn}
\end{figure*} 

\section{Final Remarks}

\indent \par In this article, we continued to develop a study on electromagnetic fields in the presence of a boosted Kerr black hole, generalizing the axial case used in \cite{aranha}. Following the standard procedure, we adopt the metric obtained in \cite{ids1} written in Robinson-Trautman coordinates. Soon after, we generalize to the cases of Bondi-Sachs and Kerr-Schild coordinates. Given the absence of axial symmetry, the electromagnetic fields obtained in this work could not be constructed via ZAMOs \cite{takahashi}. The reason is simply that the axial Killing vector $\partial / \partial \phi$ is not a Killing vector in the general case, leaving this role only for the TLKV $\partial / \partial t$. In this way we also choose a general timelike observer to describe the eletromagnetic fields in the infinity instead of the usual ZAMOs. The main point of this generalization for the non-axial case is given by the boosts in arbitrary directions. This considerably increases the electromagnetic effects found in the vicinity of the black hole, thus indicating possible and relevant astrophysical applications.
\indent \par As in the previous article, the spacetime considered is also expanded in a $1/r$ series and, up to the first order, the metric can be shown as a solution of the Einstein equations. Then we constructed the Maxwell fields via isometries up to the order considered and show that all electric field components are null differently of its magnetic counterparts. In this sense, we can say we have a case, up to this order, similar to a quasineutrality regime as seen in some plasmas. We plan to explore this in a future work.
\indent \par From a qualitative point of view, magnetic fields behave differently depending on the parameters of the black hole. In this work we focus mainly on the boost and rotation parameters. As can be seen in the several plots displayed in Section IV, the boost parameter, when increased, intensifies the distortion of the magnetic field lines. Likewise, in an analysis of the boost direction parameter (or from another perspective, the rotation direction) of the black hole, we can see graphically that the magnetic field is more intense when the rotation and the boost direction are aligned, losing power as the associated vectors begin to misalign. We observe the same structure in all coordinate system adopted (KS or BS).
\indent \par In conclusion, let us consider two points associated with future developments. The first is related to the possibility of obtaining the boosted Kerr-Newman solution. According to the work of Fayos and Sopuerta \cite{sopuerta}, the Papapetrou field for Kerr spacetime is equivalent to the Kerr-Newman field. This indicates that, even in a vacuum Kerr geometry isometries can generate a physical field not just an analogy, as described by some authors\cite{nouri-zonoz}. 
As demonstrated in this work the magnetic field generated by the intrinsic geometry has particular
features   
which are intensified depending of the boost parameter.
The second point is linked to the qualitative aspects of the Blandford-Znajek effect in the presence of a Papapetrou field of a boosted Kerr black hole. As we demonstrated in this work (and in the previous one), the magnetic fields generated by the intrinsic isometry of the system have characteristic distortions which are intensified depending on the boost parameters (velocity and direction). The alignment of the magnetic field with the rotation axis of, in addition to the direction of the boost, are very important for the description of the astrophysical scenarios which occur around the black hole. In a recent work \cite{science}, a x-ray binary source, MAXI J1820+070, has a structure which could be related to the system we constructed here. Moreover, the study of jet within a system like this one could bring several interesting results and we are looking forward to investigate such possible implications.

\end{document}